\documentclass[prl,twocolumn,showpacs,superscriptaddress]{revtex4}
\usepackage{amsmath,amssymb,graphicx,epstopdf,color,bm,soul,upgreek}

\begin{document}

\title{Exciton-Polariton Trapping and Potential Landscape Engineering }

\author{C. Schneider}
\affiliation{Technische Physik, Wilhelm-Conrad-R\"ontgen-Research Center for Complex Material Systems, Universit\"at W\"urzburg, Am Hubland, D-97074 W\"urzburg, Germany}

\author{M.D. Fraser}
\affiliation{Quantum Functional Systems Research Group, RIKEN Center for Emergent Matter Science, 2-1 Hirosawa, Wako-shi, Saitama 351-0198, Japa}

\author{K. Winkler}
\affiliation{Technische Physik, Wilhelm-Conrad-R\"ontgen-Research Center for Complex Material Systems, Universit\"at W\"urzburg, Am Hubland, D-97074 W\"urzburg, Germany}

\author{M. Kamp}
\affiliation{Technische Physik, Wilhelm-Conrad-R\"ontgen-Research Center for Complex Material Systems, Universit\"at W\"urzburg, Am Hubland, D-97074 W\"urzburg, Germany}

\author{Y. Yamamoto}
\affiliation{ImPACT Project, Japan Science and Technology Agency, Chiyoda-ku, Tokyo 102-0076, Japany}
\affiliation{Edward L. Ginzton Laboratory, Stanford University, Stanford, California 94305-4085, USA}

\author{E. Ostrovskaya}
\affiliation{Nonlinear Physics Centre, The Australian National University, Canberra ACT 2601, Australia}

\author{S. H\"ofling}
\affiliation{Technische Physik, Wilhelm-Conrad-R\"ontgen-Research Center for Complex Material Systems, Universit\"at W\"urzburg, Am Hubland, D-97074 W\"urzburg, Germany}
\affiliation{SUPA, School of Physics and Astronomy, University of St Andrews, St Andrews KY16 9SS, United Kingdom}

\begin{abstract}
Exciton-polaritons in semiconductor microcavities have advanced to become a model system for studying dynamical Bose-Einstein condensation, macroscopic coherence, many-body effects, nonclassical states of light and matter, and possibly quantum phase transitions in a solid state. Being low mass bosons, these light-matter quasiparticles can condense at comparably high temperatures up to 300K, while preserving fundamental properties such as coherence in space and time domain even when they are out of equilibrium with the environment. Although the presence of an in-plane polariton confinement potential is not strictly necessary in order to observe condensation of polaritons, engineering the polariton confinement is a key to controlling, shaping and directing the flow of polaritons. Prototype polariton-based optoelectronic devices rely on ultrafast photon-like velocities and strong nonlinearities, as well as on tailored confinement. Nanotechnology provides several pathways to achieving such a confinement, and the specific features and advantages of the different techniques are discussed in this paper. As hybrid exciton-photon quasiparticles, polaritons can be trapped via their excitonic as well as their photonic component, which leads to a wide choice of  highly complementary techniques. Here we highlight the almost free choice of trapping geometries and depths of confinement that provides a powerful tool for control and manipulation of polariton systems both in semi-classical and quantum domain. Furthermore, the possibility to observe effects of polariton blockade, Mott insulator physics, and population of higher-order bands in sophisticated lattice potentials is discussed. The observation of such effects will signify the opportunity for the realization of novel polaritonic non-classical light sources and quantum simulators.
\end{abstract}

\maketitle

\section{Introduction}

The term 'polariton' is generally used in solid state physics, when an optical excitation couples strongly to a matter excitation.  The matter excitation can be provided by a plasmon, a phonon, an electron or an exciton. This review article exclusively discusses exciton-polaritons in microcavity systems. The excitons, which are bosonic composite quasi-particles consisting of an electron and a hole bound via their Coulomb attraction in a semiconductor, can be confined in quantum well (QW) structures embedded in optical microcavities. If the conditions for strong light-matter coupling are fulfilled, the properties of the bosonic matter excitation and the photon light field inside the microcavity are mixed, and new eigenstates of the coupled system evolve \cite{Kavokin, Yamamoto}.
Being bosonic quasi-particles, polaritons can in principle condense in a single particle energy state of finite size \cite{Kasprzak-Nature06}. This dynamical condensation of bosons is closely related to the Bose-Einstein Condensate (BEC) phase which is usually studied in ultra-cold atomic systems \cite{Anderson-Science95}. However due to the finite lifetimes of polaritons stored even in state-of-the-art microcavities, thermal equilibrium between the polaritons and their environment is very hard to establish. Nevertheless, long-range spatial \cite{Kasprzak-Nature06, Balili-Science07}  and temporal \cite{Deng-Science02, Tempel-NJP12, Kasprzak-PRL08} coherence of polariton condensates has been demonstrated, revealing signatures of a BEC \cite{Deveaud-JOSAB12}.  The effective mass of a microcavity polariton is approximately 5 orders of magnitude smaller than that of a free electron and 8-9 orders smaller than that of an atom. Since the critical temperature for the Bose condensation scales inversely with the particle mass \cite{Einstein24}, polaritons are well suitable for studies of condensation in the temperature range from liquid helium up to even room temperature.

Polaritons exist in a semiconductor microcavity environment, hence their properties can be tailored by means of semiconductor lithography and nanotechnology. Tailoring polaritonic systems via spatial trapping of excitons and/or photons in real and momentum space enables to control and manipulate these quasi-particles, and therefore can open completely new areas of mesoscopic physics in semiconductors. While photons have to be confined in a cavity to realize sufficiently long lifetimes to form polaritons, the excitons which we consider in this article are usually located in quantum wells to enhance the oscillator strength of the emitters and to reduce possible detrimental effects of surface recombination.  Consequently, polaritons which are generated in the above mentioned manner are quasi-particles living in a two-dimensional environment of a quantum well-microcavity system. In order to allow for the observation of phase transitions related to Bose-Einstein Condensation at finite temperatures, the exciton-polariton system should be of finite size,  since otherwise the condensate's long range order would be destroyed by thermal fluctuations \cite{hohenberg}. This condition is usually assured by the finite area of optical (or electrical \cite{schneider2013electrically}) excitation.  Engineering of additional lateral traps in the semiconductor yields the possibility to create sophisticated potential landscapes on the order of the polariton wavelength (on the order of 1 $\mu$m), and therefore to study interactions and transport of tailored condensates in complex environments. This possibility, in turn, can be exploited in designing new schemes of highly nonlinear photonic integrated circuits (PICs) and logic elements  \cite{Liew-PRL08, Liew-PRB10}. Such PICs promise ultra high processing speeds at very low power consumption, since polaritons can propagate with ultrafast velocities and low decoherence rates in the semiconductor. The inherently high non-linearities borrowed from the excitonic components allow to manipulate \cite{Sturm-NatComs14}, switch and steer the polariton flow with very weak additional laser beams \cite{gao-PRB12} and possibly electrical contacts. Additionally, the spin degree of freedom in the polariton system can be actively exploited to add new functionalities in polariton logic devices as discussed in \cite{liew2011polaritonic}.

The possibility to engineer polariton trapping potentials has furthermore triggered the interest in using polaritonic systems to emulate complex many-body phenomena, such as the physics of high-temperature superconductors, graphene, or frustrated spin lattices \cite{Kim-NatPhys11, Jacqmin-PRL14, Masumoto-NJP12}. Quantum emulators are envisaged as a highly desirable tool for understanding complex many-body properties of novel solid state, chemical, and biological systems, which are otherwise hardly accessible. They rely on the emulation of Hamiltonians via potential landscape engineering in a highly controllable quantum system \cite{Cirac-NatPhys12}. Polariton gases in microcavities are considered as promising candidates for solid state quantum emulation, as they fulfil a range of important prerequisites: They can form bosonic condensates and enter a superfluid phase \cite{Carusotto-RMP13} they possess internal (pseudo-spin) degrees of freedom, and as we will further assess in great detail, they can be localized by lithographic or optical techniques possibly down to the single polariton level, and their interaction constants are tunable \cite{Takemura-arxiv13, Takemura-NatPhys14}.
One example of progress towards quantum emulation are periodic potential landscapes, where higher-band p- and d-orbital like condensates mimic the fundamental structure of high temperature superconductors \cite{Lai-Nature07, Kim-NatPhys11}, offer the possibility to study Dirac cones \cite{ Kim-NJP13,Jacqmin-PRL14}, and are predicted to implement Bose-Hubbard Hamiltonian physics in semiconductors \cite{byrnes2010mott}.

Predictions of the quantum blockade regime in dilute polariton systems have gained significant attention \cite{Verger-PRB06, Liew-PRL10, Bamba-PRA11, besga-arxiv13}. This effect would allow to exploit the fascinating properties of polaritons in integrated quantum light sources, sources of entangled and indistinguishable photons \cite{Na-NJP10}, and to generate polariton number states in microcavity traps. The latter effect could pave the way for the study of quantum phase transitions and Bose-Hubbard physics with light,  which yield the potential to open new areas in the field of solid state microcavity research.  However, the experimental demonstration of polariton quantum blockade critically relies on the tight trapping of polaritons to enhance polariton-polariton interactions \cite{Verger-PRB06} or controlled coupling of polariton boxes to facilitate quantum interference \cite{Bamba-PRA11}. The experimental and technological difficulties are significant, and consequently the effect has not been demonstrated yet.

The rapidly expanding research on exciton-polaritons creates an unrelenting demand for elegant and non-destructive methods for trapping polaritons in a microcavity. This articles summarizes a number of complementary techniques that have been developed to meet the demands. We discuss techniques to trap the excitonic component of the polaritons by manipulating the QWs or structuring the optical pump, as well as methods for trapping the photonic component by manipulating the cavity.  We review the most prominent experiments enabled by each of the techniques and discuss the limitations and prospects.

\section{Theory}

\subsection{Quantum well excitons}

The excitons are localized in the plane of the QWs, however in-plane the exciton wavefunction is usually delocalized over a large number of the crystalline lattice sites due to the electromagnetic screening in the semiconductor with a characteristic Bohr radius of 5-15 nm and the binding energy of 4-10 meV for typical III/V compounds (except nitrites). Since the physics of quantum well excitons has been studied for several decades, here we will only summarize the most important physical relations and parameters and refer the interested reader to the exhaustive literature (see, e.g. \cite{Klingshirn, Kavokin, weisbuch, Masselink-PRB85}).


The energy spectrum of a Wannier-Mott exciton in a crystal with a dielectric constant $\epsilon$ can be found in the effective mass approximation, by solving a Schr\"odinger's equation for an electron in a  hydrogen atom, where the electron mass has to be replaced by the reciprocal mass $\mu=\frac{m_e m_h}{m_e+m_h}$ and the dielectric constant $\epsilon_0\rightarrow \epsilon \epsilon_0$.

The Schr\"odinger equation for the relative electron-hole motion
\begin{equation}
(\frac{\hbar^2 \nabla^2}{2 \mu}+\frac{e^2}{4\pi\epsilon\epsilon_0 r})\psi(r)=E \psi(r)
\end{equation}
where r is the relative distance between the electron and the hole, has the solution in the form of a  the 1s scattering wavefunction
\begin{equation}
\psi(r)=\frac{1}{\sqrt{\pi a_B^3}}exp[-\frac{r}{a_B}]
\end{equation}
with the Bohr radius $a_{B}=\frac{4\pi\epsilon\epsilon_0\hbar^2}{\mu e^2}$ and the corresponding eigenenergy
$\frac{2 \mu e^4}{\hbar^2 (8 \pi \epsilon \epsilon_0)^2}$.

In quantum confined systems, such as a quantum well, the reduced dimensionality leads to modifications in these quantities, and the Bohr radius, as well as the binding energies of the excitons, critically depend on material parameters. In the presence of confinement, excitonic effects generally are enhanced, resulting e.g. in a four fold increase of the exciton binding energy in an ideal two-dimensional system. The effects of confinement on excitons are discussed in the literature, e.g. \cite{Kavokin, Andreani-PRB90, Klingshirn, Masselink-PRB85} to list a few.

The oscillator strength of QWs can be expressed in terms of the overlap integral between the electron and hole wavefunctions. Excitonic effects lead to an increase of the oscillator strength compared to band-to-band transitions. The oscillator strength for a transition $f$ can be expressed in the effective mass approximation in terms of the lattice periodic (Bloch) part of the wavefunctions in the valence (v) and conduction band (c) $u_{v,c}$ and the envelope $F_{v,c}$ with the momentum operator $\bf p$ and the polarization vector $\hat{e}$:
\begin{equation}
f=\frac{2}{\mu \hbar \omega }|\left\langle u_v|\hat{e} {\bf p}|u_c   \right\rangle|^2|\left\langle   F_v|F_c\right\rangle|^2
\end {equation}

Here, $\hbar\omega$ is the energy of the considered transition.

The oscillator strength is closely related to the optical absorption in a QW via
\begin{equation}
A=\frac{4\pi^2e^2\hbar f}{n m_0 c L},
\end{equation}
it can be determined via photon absorption spectroscopy \cite{Masselink-PRB85}. Here n is the refractive index of the material, $m_0$ the free electron mass and L the the well width. A is the integrated absorption intensity (in units of $\frac{eV}{cm}$).

\subsection{Microcavity Photons}

\begin{figure}[ht]
\centering
\includegraphics[width=7cm]{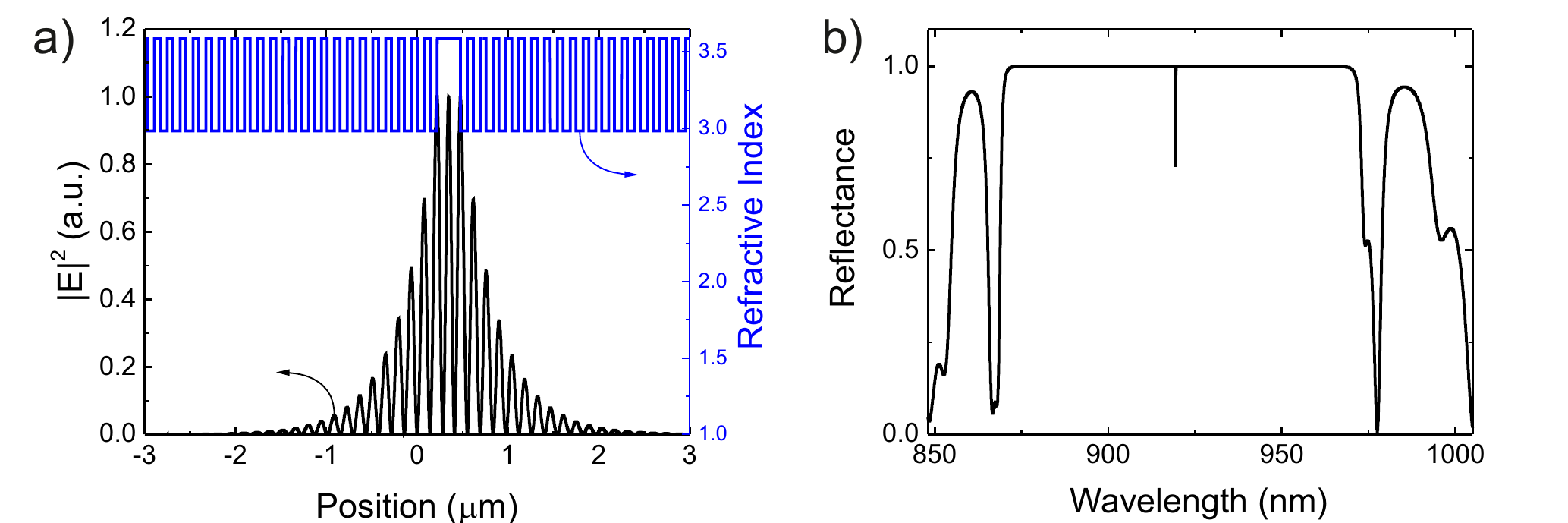}
\caption{\label{Abb:Fig2} a) Vertical mode profile in the microcavty: The optical field is strongly enhanced in the central layer, where the QWs are located to maximize light-matter coupling. b) Calculated reflectivity spectrum of a DBR-microcavity with a strongly pronounced Fabry-Perot resonance.
}
\end{figure}

There are a wide range of possibilities to trap photons in semiconductor microcavities, in one or more dimensions. Common techniques involve total internal reflection on semiconductor-air interfaces, photonic bandgap tailoring, plasmonic resonances or distributed Bragg reflection \cite{Kavokin}. In the field of microcavity polaritons, distributed Bragg reflector (DBR) based microcavities are most commonly used, since they can provide near unity reflectivity without additional sample processing and lithography steps. We will briefly summarize their basic properties in this section.

The typical cavity   hosting exciton-polaritons is composed of layers with different refractive indices, such as AlAs and (Al)GaAs. If the thicknesses of the individual layers are chosen to match the Bragg condition, $d=\lambda/(4n)$, then in such a DBR, interference of transmitted and reflected light allows to achieve extraordinary high reflectivity in a spectral window with a width of about 100 nm around the Bragg wavelength.
In order to form a cavity based on such reflectors,  a photonic defect layer has to be included. We will only consider the simplest case here, namely a $\lambda/n$- thick defect of a high-index material (such as GaAs).

The optical mode profile in such a Fabry-Perot resonator can be calculated by the transfer-matrix method, yielding a strong enhancement of the light intensity inside the cavity defect layer and an oscillating decay inside the DBR mirrors. Details of the method are described in \cite{Kavokin}. From this decay, we can define an effective cavity length, which is the sum of the physical cavity length $L_c$ and the penetration depth of the light into the mirrors:  $L_{\rm eff}=L_c+\frac{\lambda_c}{2n_c}\frac{n_1n_2}{|n_1-n_2|}$, where $n_i$ are the refraction indices of the DBR layers and $n_c$ is the refractive index of the cavity layer.

Fig. \ref{Abb:Fig2}a) depicts the optical field inside of the cavity, as calculated by the plane-wave expansion technique. Additionally, Fig. \ref{Abb:Fig2}b) shows the reflectivity spectrum of the full structure consisting of the microcavity and the surrounding DBRs, which is characterized by a high-reflectivity region (the stopp-band) and the centered resonance dip. The width of this resonance defines the so-called quality factor or Q-factor via $Q=\omega/\delta \omega$, which is directly linked to the lifetime $\tau$ of photons in the cavity $Q=\omega \tau$. In (Al)GaAs DBR based microcavities, Q-factors of several 100000 are available \cite{Reitzenstein-APL07, Schneider-arxiv2015} which reflects the high material qualities available by state-of-the-art epitaxy.

As the photon cannot freely propagate through the microcavity, its dispersion relation is strongly modified, and it can be approximated by a parabola for small in-plane momenta, $k_{||}$, which results in an effective photon mass in a microcavity:
\begin{equation}
E(k)\approx E_0+\frac{\hbar^2}{2m_c} k_{||}^2
\end{equation}
This effective mass is defined as $m_c=\frac{hn_c}{cL_c}$ and has a value on the order of $10^{-5} m_0$.  A more detailed derivation of this equation can be found e.g. in \cite{Deng2010}.

\subsection{Exciton-polaritons and normal mode coupling}

Implementing a QW with a high oscillator strength in a microcavity with a sufficiently large Q-factor can give rise to a normal mode coupling  between the photonic and excitonic resonance. The energy is transferred back and forth from the excitons in the QW to the microcavity photons, which results in a characteristic Rabi-oscillation in the time domain accompanied by a normal mode splitting of the polariton branches. A semi-classical treatment of this effect by modeling the system as two damped coupled oscillators is described in \cite{Savona-SSC95, Khitrova-RMP99}. This coupled oscillator model yields an analytical expression for the normal mode coupling in a straight forward manner. Here, we only summarize some important conclusions from this model:
Solving the transfer matrix equations yields an expression which connects the frequency of the polariton states with the frequerncy of the exciton $\omega_x$, the cavity resonance $\omega_c$, the interaction potential V and the dissipative channels $\gamma_x,c$:
\begin{equation}\label{Hlin1}
(\omega_c-\omega-i\gamma_c)(\omega_x-\omega-i\gamma_x)=V^2
\end{equation}

This expression yield an equation for $\omega$, which reads

\begin{equation}\label{Hlin}
\omega=\frac{\omega_c+\omega_x-i(\gamma_x+\gamma_c)}{2} \pm \sqrt{V^2+\frac{1}{4}(\omega_x-\omega_c-i(\gamma_x-\gamma_c))^2)}
\end{equation}

\begin{figure}[ht]
\centering
\includegraphics[width=7cm]{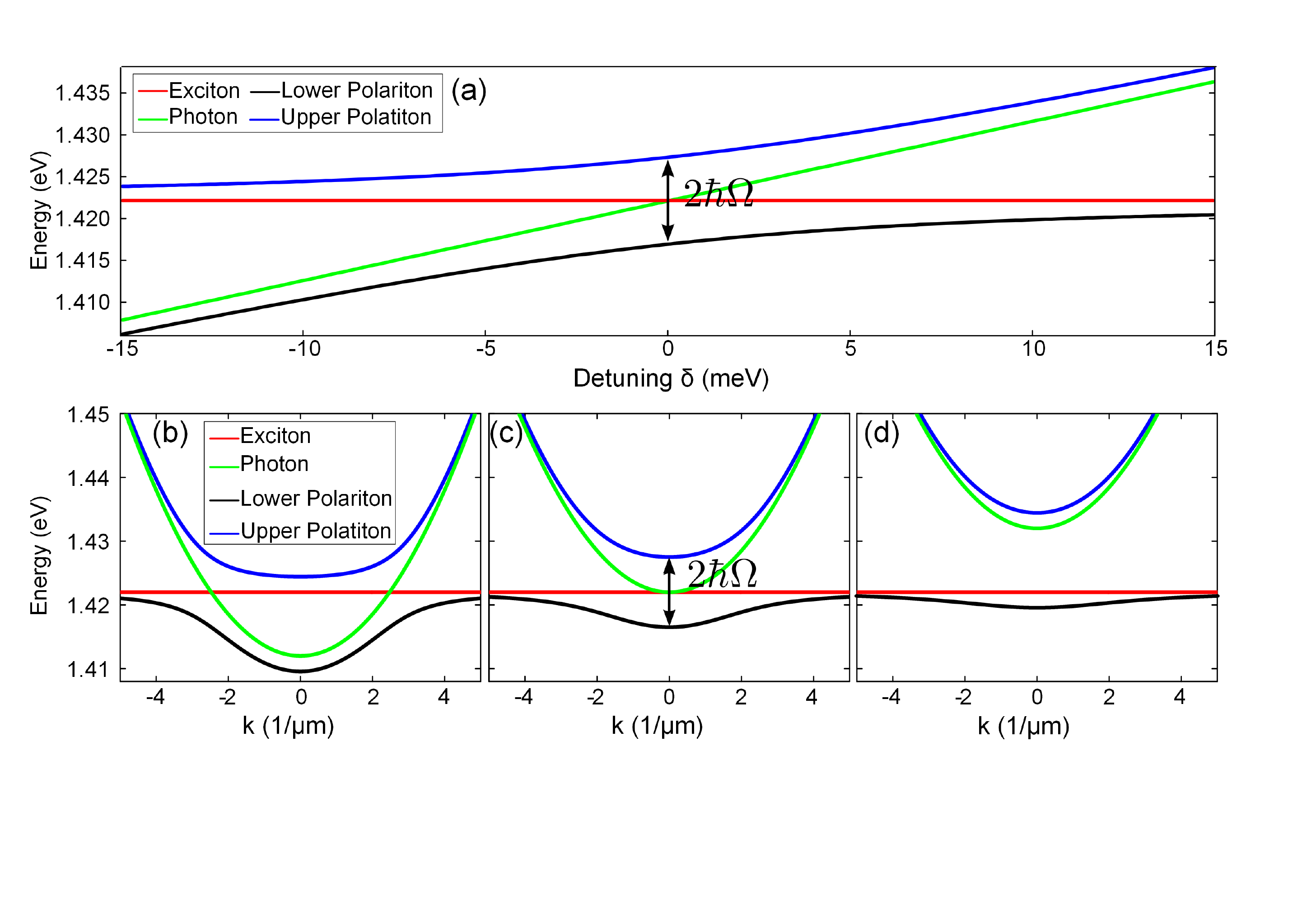}
\caption{\label{Abb:Fig3} (a) Anticrossing of the lower and upper polariton branch at zero in-plane momentum. Energy-momentum dispersion relations for exciton-polaritons with a large negative detuning of $-\hbar\Omega$ (b), at zero-detuning (c), and with a large positive detuning of $\hbar\Omega$ (d).
}
\end{figure}

The corresponding energies of the upper and lower polartion branch $\hbar\omega$ is plotted as a function of the detuning $\hbar\omega_c-\hbar\omega_x$ in Fig.~\ref{Abb:Fig3}a).

The splitting between of the upper and lower polariton branches on resonance is the characteristic Rabi splitting $\hbar\Omega$. The Rabi-frequency $\Omega$ is then defined as:
\begin{equation}
\Omega=2\sqrt{V^2-\frac{1}{4}(\delta \omega_c - \omega_x)}
\end{equation}

The coupling constant V is a function of the exciton oscillator strength, the number of QWs in the microcavity and the effective cavity length $L_{eff}$:

\begin{equation}
V\approx \sqrt{\frac{2 \Gamma_0 c N_{QW}}{n_c L_{eff}}}
\end{equation}

with
\begin{equation}
\hbar\Gamma_0=\frac{\pi e^2 \hbar}{n_c 4\pi \epsilon_0 m_e c} f_{n}
\end{equation}

The hybridization of light and matter leads to a strong modification of the energy-momentum dispersion relations. In order to derive the polariton dispersion relations, one needs to introduce the Hopfield coefficients $|X|^2$ and $|C|^2$ which determine the fraction of light ($|C|^2$) and matter ($|X|^2$) in the polaritons:
\begin{equation}
|X|^2=0.5\left[1+\frac{E_x-E_c}{\sqrt{\hbar^2\Omega^2+(E_x-E_c)^2}}\right]
\end{equation}
and
\begin{equation}
|C|^2=0.5\left[1-\frac{E_x-E_c}{\sqrt{\hbar^2\Omega^2+(E_x-E_c)^2}}\right]
\end{equation}
As a result of the parabolic cavity dispersion, the lower polariton dispersion also acquires a parabolic shape around its energy ground state, and the effective mass of the lower polariton critically depends on the exciton photon detuning, $\delta=E_x-E_c$, through the Hopfield coefficients: $m^{-1}_{LP}=|X|^2m^{-1}_x+|C|^2m^{-1}_c$ (see. Fig~\ref{Abb:Fig3})

\subsection{Polariton lasing and condensation}

\begin{figure}[ht]
\centering
\includegraphics[width=7cm]{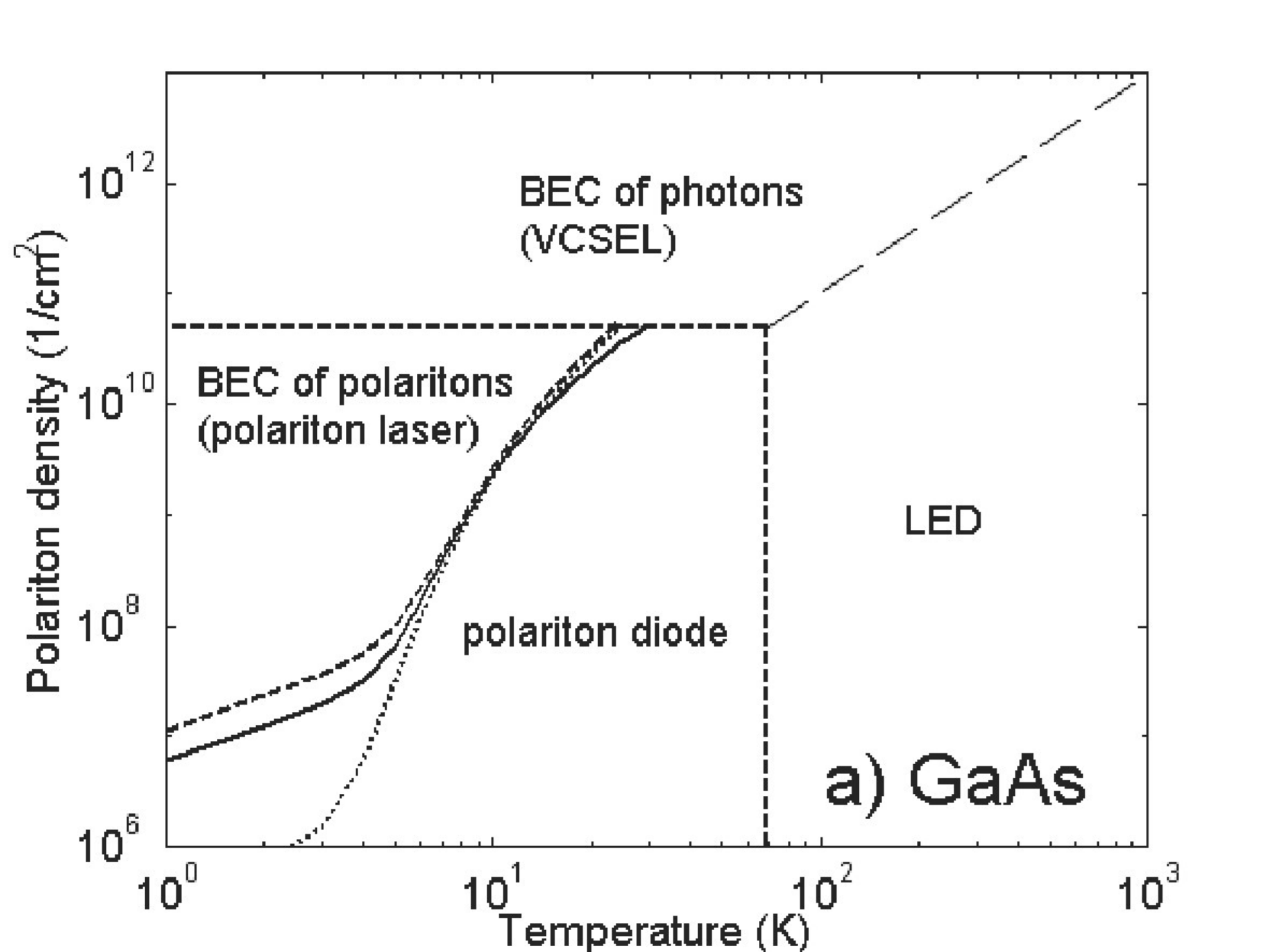}
\caption{\label{Abb:Fig4} Phase diagram of exciton-polaritons based on GaAs QWs in a microcavity. The image is reproduced from \cite{kavokin2003semiconductor}.
}
\end{figure}

Exciton-polaritons behave as bosons in the low density limit. Due to their light effective mass they are excellent candidates for studies of bosonic condensation effects and related phenomena at elevated temperatures in semiconductors. However, there are several restrictions:

\begin{itemize}
\item Polaritons only live for a short time (1-100 ps) in a microcavity before they decay radiatively. Hence, they can hardly reach true thermal equilibrium. Therefore, the condensate of exciton-polaritons is often referred to as a dynamic, open-dissipative condensate, or a polartion laser. A comprehensive discussion of the 'polariton-laser' vs 'condensate' can be found e.g. in \cite{byrnes2014exciton}.
\item At high densities, screening of excitons is dominant, which leads to a bleaching of the oscillator strength and the system enters the (fermionic) plasma phase. This typically happens for particle densities on the order of the Mott transition, i.e $10^{10}-10^{11}$ electron-hole pairs per $cm^{-2}$ \cite{houdre1995saturation}.
   \item Excitons are only stable up to a certain temperature which is determined by their binding energy.
\end{itemize}
 The resulting phase diagram of exciton-polaritons has been drawn by Kavokin et al. \cite{Kavokin}, and is sketched in fig.~\ref{Abb:Fig4} for the case of GaAs QWs in a microcavity.

The planar microcavity exciton-polariton system is inherently two-dimensional (2D), and in accordance with the Hohenberg-Mermin-Wagner theorem, the transition to BEC in a uniform system is only possible at zero temperature \cite{hohenberg} for 2D as well as 1D geometries. Restriction of the system to a finite size, however, inhibits excitation of density and phase fluctuations permitting the formation of a condensate or quasi-condensate phase with a macroscopic coherence length. This is true for a polariton system with a finite excitation area, both in one and two dimensions \cite{petrov-PRL00a, petrov-PRL00b}. In most cases, the coherence length of the polariton condensate is thus finite, and ranges on the order of the optical pumping spot size~\cite{Kasprzak-Nature06}. At macroscopic lengths, and in the 2D case, it is generally assumed that the system enters the Berezinskii-Kosterlitz-Thouless (BKT) phase, where vortex-antivortex pairs are formed, and the spatial correlation function decays according to a power law \cite{berezinskii1972destruction, kosterlitz1973ordering}. Although the correlation function measurements in exciton-polariton system are suggestive of the BKT phase \cite{nitsche2014algebraic} more conclusive detection of spontaneously formed vortex pairs is required to confirm this regime.

The need to confine polaritons goes far beyond the unambiguous observation of condensation effects. As described in the introduction, the engineered confinement of exciton-polaritons paves the way for creating functional polaritonic circuits and quantum simulators, as well as fundamental studies of polariton condensation in complex potential landscapes.

\subsection{Polaritons in potential landscapes}
\label{bandstructure}
Since exciton-polaritons are composed of light and matter, both the photonics and the excitonic part of the quasiparticle can be subject to confinement. Theoretical description of the exciton-polaritons in the engineered potentials below condensation threshold generally relies on calculating the eigenstates of photons and excitons under respective confinement conditions and then using the exciton-photon interaction Hamiltonian to derive the corresponding polariton dispersion. This approach was successfully used, e.g., to describe discrete polariton states appearing in a zero-dimensional 'mesa' trap \cite{Kaitouni-PRB06}. Indeed, as  will be discussed below in sec. \ref{etch-and-overgrowth}, such a trap provides a purely photonic confinement of a polariton, and therefore it is sufficient to solve a Maxwell equation for the photon field, ${\mathcal E}({\bf \rho})$ with the boundary conditions given by the contour of the mesa, and assuming that at a position ${\bf \rho}$ within the trap the electromagnetic modes are locally equivalent to the planar cavity modes: ${\mathcal E}(r)={\mathcal E}({\bf \rho}) \exp(ik_z({\bf \rho})z)$, where $k_z(\bf{\rho})$ is a suitably modified propagation constant. Introducing the field operators of cavity photons, $\hat{\phi}$, and excitons, $\hat{\chi}$, the second-quantized Hamiltonian for the two coupled oscillators model (\ref{Hlin}) \cite{Deng2010} can be written as:
\begin{equation}
\begin{array}{l}
\hat{H}=\sum_{\{n\}}\left(\hbar \omega_{c\{n\}}\hat{\phi}_{\{n\}}^\dagger\hat{\phi}_{\{n\}}+\hbar \omega_{x\{n\}}\hat{\chi}_{\{n\}}^\dagger\hat{\chi}_{\{n\}}\right )+\\
\sum_{\{n\},\{n'\}}\left(\hbar \Omega_{\{n\},\{n'\}} \hat{\phi}_{\{n\}}^\dagger \hat{\chi}_{\{n'\}}+{\rm h.c.}\right),
\end{array}
\label{eq1}
\end{equation}
where summation over all quantum numbers of the bound states is assumed. Here $\hbar\omega_{c,x}$ are the eigenenergies of the photon and exciton modes, and $\Omega_{\{n\},\{n'\}}$ is defined through the Rabi splitting in the planar region, $\Omega$, and the overlap integrals between the corresponding exciton and photon modes \cite{Kaitouni-PRB06}. In the absence of excitonic confinement, exciton modes are those of a free particle.

\begin{figure}[ht]
\centering
\includegraphics[width=8cm]{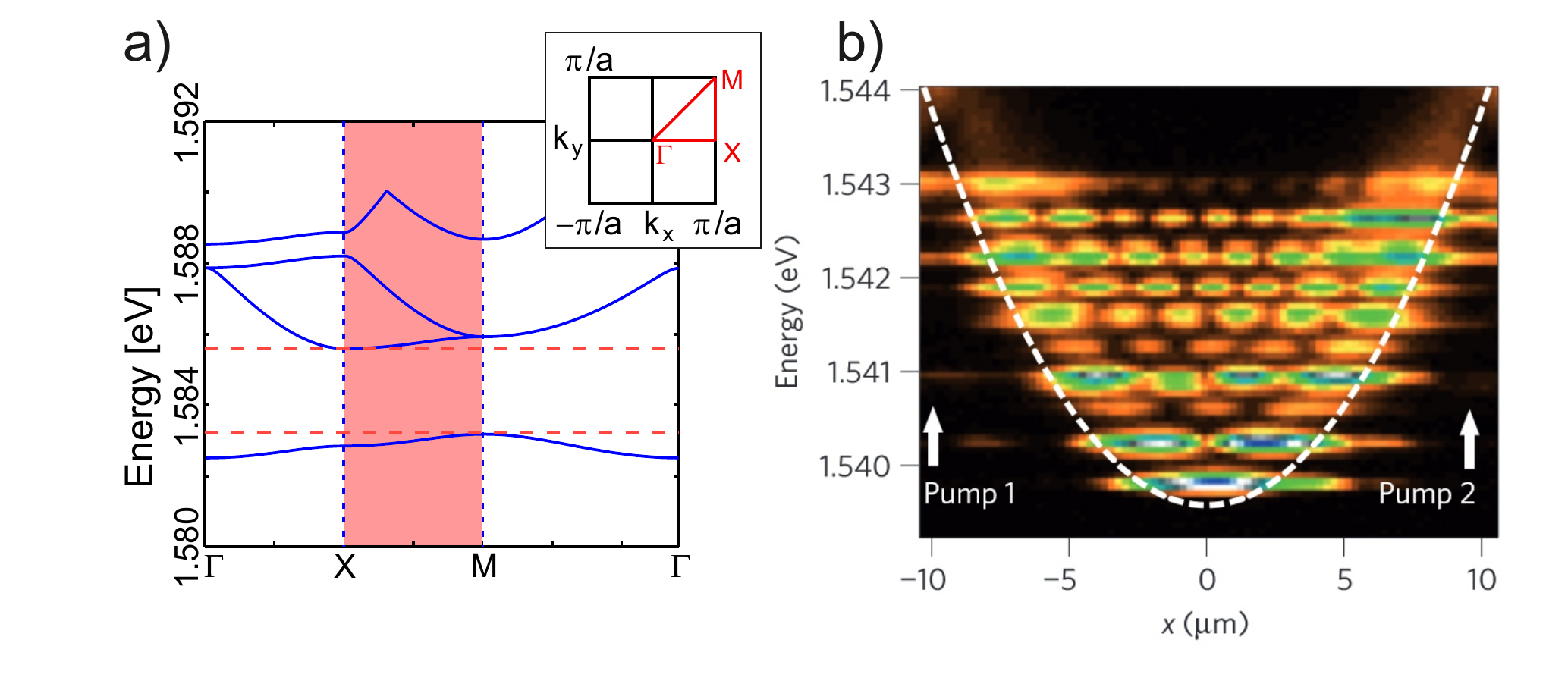}
\caption{\label{Abb:Bandstructure} a) The first four energy bands for a square lattice
array of polariton traps in the reduced Brillouin Zone (BZ) representation. Dashed lines indicate the position of the complete band gap. Inset: The first irreducible BZ. b) Harmonic energy spectrum of a polariton condensate trapped between two laser spots. This figure is reproduced with permission from \cite{Tosi-NatPhys12}.
}
\end{figure}

Similarly, using the linear exciton-photon coupling Hamiltonian (eq.~\ref{Hlin}, eq.~\ref{eq1}), it was shown that the structure of the energy bands imposed either on the cavity photon mode (or the exciton mode) by a periodic potential (e.g., periodic array of mesas) translates into the band-gap structure of the polariton spectrum \cite{Boiko2008, Winkler-NJP15}. For the case when only photonic modes are affected by the periodic potential, the energy bands of the lower-polariton (LP) state are determined as:
 \begin{equation}
E_n^{LP}=\frac{1}{2} [E_{x}+E_{cn}-\sqrt{\hbar^2\Omega^2+(E_{cn}-E_{x} )^2} ],
\end{equation}
where $E_{cn}=\hbar \omega_c$ are the photonic Bloch bands, and $E_x=\hbar\omega_x$. Alternatively and equivalently, the LP band structure can be calculated directly by solving the eigenvalue problem for the polariton Bloch states $u_k ({\bm r})=u_k ({\bm r}+{\bf a})$  in an {\em effective} polaritonic in-plane potential $V(\bm r)=V({\bm r}+{\bf a})$:
\begin{equation}
\left[\frac{\hbar^2}{2m_{LP}} (-i\nabla+{\bf k})^2+V({\bm r})\right] u_{(n,{\bf k})} (\bm r)=E_n ({\bf k}) u_{(n,{\bf k})} ({\bf r})
\end{equation}
where n is the band index and $m_{LP}$ the effective polariton mass in the planar region. A typical structure of LP energy bands $E_{LP}({\bf k})$ is plotted in Fig.~\ref{Abb:Bandstructure} for a square lattice with $a=3$ $\mu$m, and the comparison with the low density polariton emission spectrum (see sec. \ref{etch-and-overgrowth}, Fig. \ref{fig:fig4}) demonstrates a good agreement. Furthermore, provided the effective polariton potential is deduced with sufficient accuracy, a fractal band-gap spectrum in a Fibbonacci lattice has also shown to be well reproduced by this approach \cite{tanese2014fractal}.

Above the condensation threshold the linear coupled oscillator model no longer holds. With the exception of a very strong confinement in zero-dimensional structures, static and dynamic features of the exciton-polaritons condensates in trapping potentials can be well captured by the mean-field description. The mean-field models are derived under the assumption that a significant population is present in each of the intracavity modes, as well as near the energy minimum of the LP dispersion, such that bosonic stimulated scattering into the condensed state is notable.

In the case of coherent, resonant excitation of the condensed state, such as the OPO regime \cite{Carusotto2013}, the modified Gross-Pitaevskii equation for the cavity photon and  QW exciton mode amplitudes can be written in the form:
\begin{equation}
\begin{array}{l}
i{\partial _t}{\Psi_c} = \left(\omega^0_c-\frac{\hbar^2}{2 m_c}\nabla^2 -i\gamma_c  \right){\Psi_c}  +\frac{\Omega}{2}{\Psi_x  } +\alpha {\mathcal E}_p \\
i{\partial _t}{\Psi_x  } =g\Psi_x+ \left( i\omega^0_x +V({\bf r},t)-i\gamma_x \right){\Psi_x  } +G|\Psi_x|^2{\Psi_x}.
\end{array}
\label{eq2}
\end{equation}
 where ${\mathcal E}_p({\bf r}, t)={\mathcal E}_0\exp(i{\bf k} {\bf r})\exp(-i\omega_pt)$ is the optical pump field, $\alpha$ is the response coefficient, $G$ is the exciton-exciton interaction strength, and the exciton and photon modes are characterized by their minimum energies $\hbar\omega^0_{c,x}$. The model in this form was successfully used, e.g., in \cite{Krizhanovskii2013} to describe dynamics of condensation in a periodic potential $V({\bf r},t)$ imposed predominantly onto excitonic component of polaritons by surface acoustic wave modulation. Note that the effective mass of the excitons is much larger than that of cavity photons, so that their kinetic energy is ignored in eq.~(\ref{eq2}).

 The model (eq.~\ref{eq2}), with the external potential term introduced into the equation for the cavity photon amplitude, could also serve to describe resonant excitation and nonlinear dynamics of exciton-polaritons in photonics potentials, e.g., those  provided by tapered microcavities \cite{gao-PRB12}. It is important to note, however, that in order to describe realistic experimental situations, the mean-field models often need to be augmented to account for non-radiative energy relaxation of excitons. This is usually done by introducing a phenomenological damping term into a modified Gross-Pitaevskii equation \cite{wouters-PRB10} in the spirit of thermal relaxation models for atomic condensates \cite{penckwitt2002}. Recently, a stochastic Gross-Pitaevskii equation for the polariton order parameter $\Psi_{LP}$  was derived from the full microscopic theory, under the assumption of coherent pumping and phonon-assistend energy relaxation \cite{Savenko-PRL13}, and shown to be successful in describing thermal relaxation of a polariton condensate in an imposed potential landscape.

 Finally, as will be discussed in detail in subsequent sections \ref{optical-traps}, incoherent, far-off resonant excitation of a polariton condensate provides a unique opportunity to create reconfigurable traps for polaritons induced by an optical pump via an excitonic reservoir. The model for the reservoir-coupled condensate is discussed e.g. in \cite{Wouters-PRL07}. In a more general form, this open-dissipative Gross-Pitaevskii equation (or modified complex Ginzburg-Landau equation), takes the form:
\begin{equation}
\begin{array}{l}
i\hbar\frac{\partial \Psi_{LP}}{\partial t}=-\frac{\hbar^2}{2m_{LP}}\nabla_\perp^2+g_c|\Psi_{LP}|^2+g_Rn_R({\bf r},t)+\\
i\frac{\hbar}{2}(Rn_R({\bf r},t)-\gamma)\Psi_{LP},  \nonumber \\
\frac{\partial n_R}{\partial t}=-(\gamma_x+R|\Psi_{LP}|^2)n_R({\bf r},t)+P({\bf r}). 
\end{array}
\label{model}
\end{equation}
Here $\Psi_{LP}$ is the condensate wavefunction, $n_R$ is the reservoir density, and $P(\bm \vec{r})$ is the spatially modulated optical pumping rate. The critical parameters defining the condensate dynamics are the loss rates of the polaritons $\gamma$ and reservoir excitons $\gamma_{x}$, the stimulated scattering rate $R$, and the strengths of polariton-polariton, $g$, and polariton-reservoir exciton, $g_R$, interactions. Even without an engineered confining potential, the polaritons can be spatially localized in the vicinity of a single pump spot due to the gain-induced self-trapping effect, which is generic for open-dissipative systems \cite{Roumpos2010, ostrovskaya2012, ge2013pattern}. Combination of several strong pump spots or pumping with structured light beams creates an effective polariton trap due to a strong repulsive potential induced by the reservoir density distribution $n_R({\bf r})$. A striking effect of this all-optical trapping scheme is shown in fig.~\ref{Abb:Bandstructure}b) where the condensate is confined in an effective 1D harmonic potential created by two intense pump spots. The model (\ref{model}) with various modifications accounting for the adiabatic evolution of the reservoir [see, e.g., Berloff and Keeling,\cite{Keeling-PRL08}], phenomenological energy relaxation, and reservoir diffusion, is successfully used to describe experimental observations of all-optically trapped polariton condensates \cite{Tosi-NatPhys12, Cristofolini-PRL13, Askitopoulos-PRB13, Dall-PRL2014}].

Theoretical and numerical studies of the model (eq. \ref{model}) have also resulted in a number of predictions concerning the behaviour of far off-resonantly excited polariton condensates in the presence of an engineered trapping potential. For example, trap-induced deformations of the polariton density \cite{Keeling-PRL08, ostrovskaya2012}, instabilities promoting vortex formation \cite{Keeling-PRL08, Borgh-PRB12} , and detailed structure of excited states in harmonic potentials \cite{Trallero-PRB14} were described theoretically, but still await to be systematically tested in experiments.

Finally, it has been predicted that in the presence of spatial confinement, polariton interactions can be notably enhanced \cite{Verger-PRB06}. The nonlinear coefficient describing the polariton interactions in the presence of optical confinement is calculated as a function of the structure parameters in \cite{Verger-PRB06} and reads $\omega_nl=\kappa *2.67/(2R)^2$ in the presence of cylindrical confinement. $\kappa$ corresponds to the polariton blueshift without lateral confinement, and ranges on the order of $\hbar\kappa= 1.5 10^-2(\mu m)^2 meV$. If the value for $\omega_nl$ exceeds the polariton linewidth, only one polariton can be injected into the system by a resonant laser, and the regime of polariton blockade is reached.


\section {Experimental approaches to polariton confinement}

In the following, we will  discuss various techniques employed to spatially confine polaritons and highlight the advantages of each technique together with its limitations. We also summarize  important experiments made possible by using the techniques under consideration.  We will first discuss techniques to confine polaritons via their excitonic part, which is mainly carried out by locally manipulating the QW. In the second half of this section we will review commonly applied techniques which are exploited to trap the photonic part of the polaritons.

\subsection{Trapping polaritons via the excitonic part}

For all applied techniques to localize microcavity polaritons via their excitonic part, it is of crucial importance to retain a high quantum efficiency, prevent surface recombination as well as preserve a large exciton oscillator strength. These design rules naturally exclude some techniques (such as defining small quantum dots (QDs) via lithography and dry etching of a QW), and set intrinsic limits to others (e.g., application of large local electric fields).

\subsubsection{Application of local strain}

\begin{figure}[ht]
\centering
\includegraphics[width=8cm]{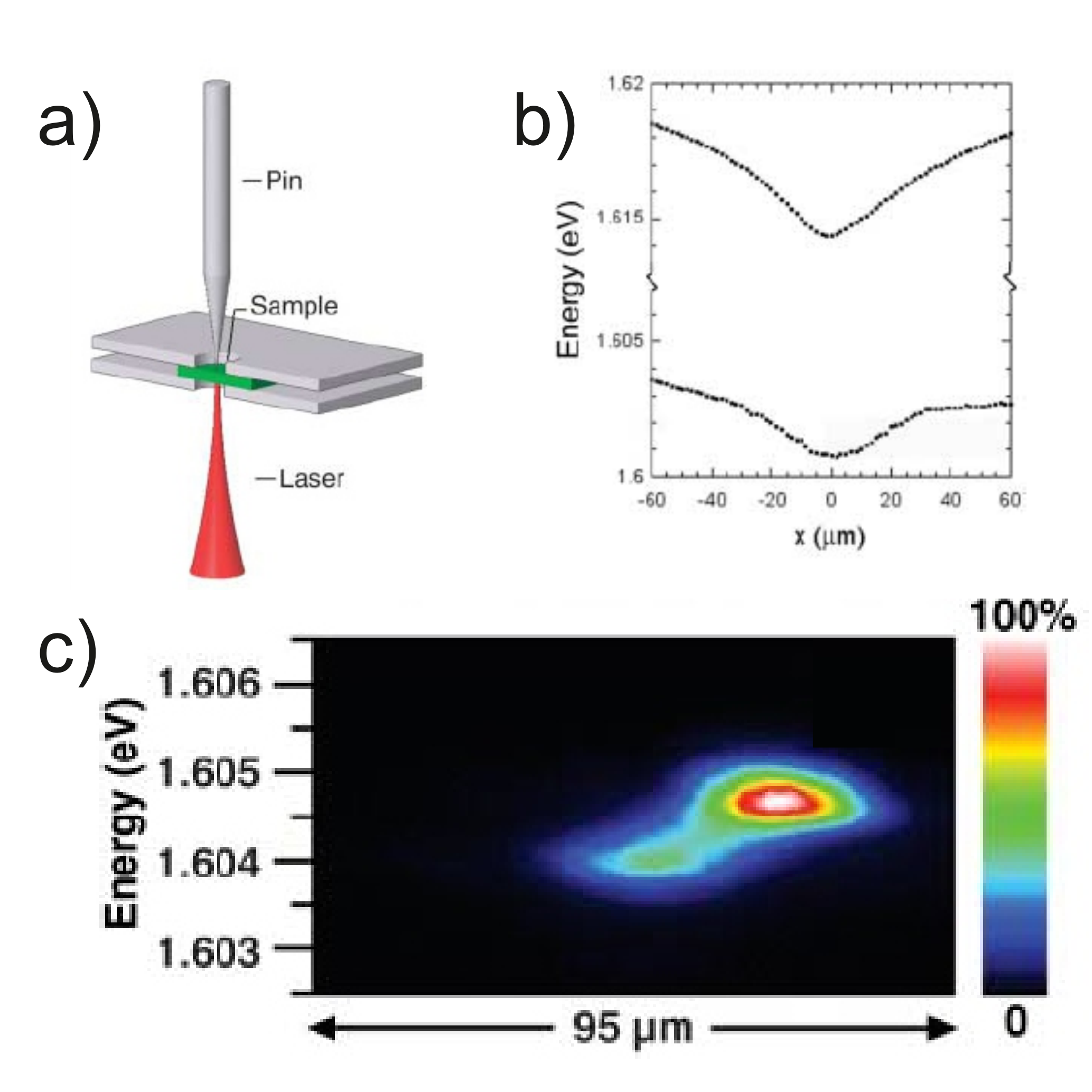}
\caption{\label{fig:Fig5} a) Schematic drawing of the setup to apply local strain to a sample via a pin. b) Potential profile on polaritons trapped in such a strained environment. c) Flow of a polariton condensate from the excitation spot (on the right) into the potential minimum created by local strain. The figure is reproduced with permission from \cite{Balili-Science07}.
}
\end{figure}

Since the bandgap of semiconductor crystals critically depends on the distance between two nearest atomic neighbors, externally applied crystal strain can be employed to locally tune the exciton wavelength. One possible scheme exploiting this effect to create a trapping potential for polaritons is presented in (Fig.~\ref{fig:Fig5}a) \cite{Balili-Science07}. The sample with a partly removed substrate is held in the cryostat, and a sharp metal tip with a radius of 50 $\mu$m locally applies compressive strain on the reverse side. In this manner, a local parabolic potential minimum for QW excitons on the order of the width of the tip can be created. The local redshift of the exciton energy at the position of the tip is also reflected in the energy minimum for exciton polaritons which is shown in Fig. \ref{fig:Fig5}b). The depth of this potential can reach several tens of  meV \cite{Balili-Science07, Balili-APL06, Voros-PRL06, Negoita-PRB99}, clearly exceeding the thermal energy at liquid helium and even nitrogen temperatures.
The implementation of this technique led to the first successful demonstration of Bose-Einstein  condensation of exciton-polaritons in a GaAs based microcavity under strictly non-resonant and non-local pumping, taking advantage of the diffusion of polaritons into the trap. This effect can be seen in the real space image in Fig. \ref{fig:Fig5}c). The polaritons are generated by a non-resonant pump laser on the right side of the trap. The emission shown in Fig. \ref{fig:Fig5}c) is a consequence of polaritons flowing into the potential minimum separated by $\sim 20$ microns from the excitation. The flexibility to adjust the strain reversibly, and hence control the potential shape and depth is another advantage of this technique. However, due to the finite size of the mechanical pin,  more complex potential landscapes beyond a single trap seem difficult to  realize. The lateral trap size on the 10-100 $\mu m$ scale makes this technique, at least in this implementation, unsuitable for the observation of size quantization effects or blockade effects on the single polariton level.

\subsubsection{Surface acoustic waves}

\begin{figure}[ht]
\centering
\includegraphics[width=8cm]{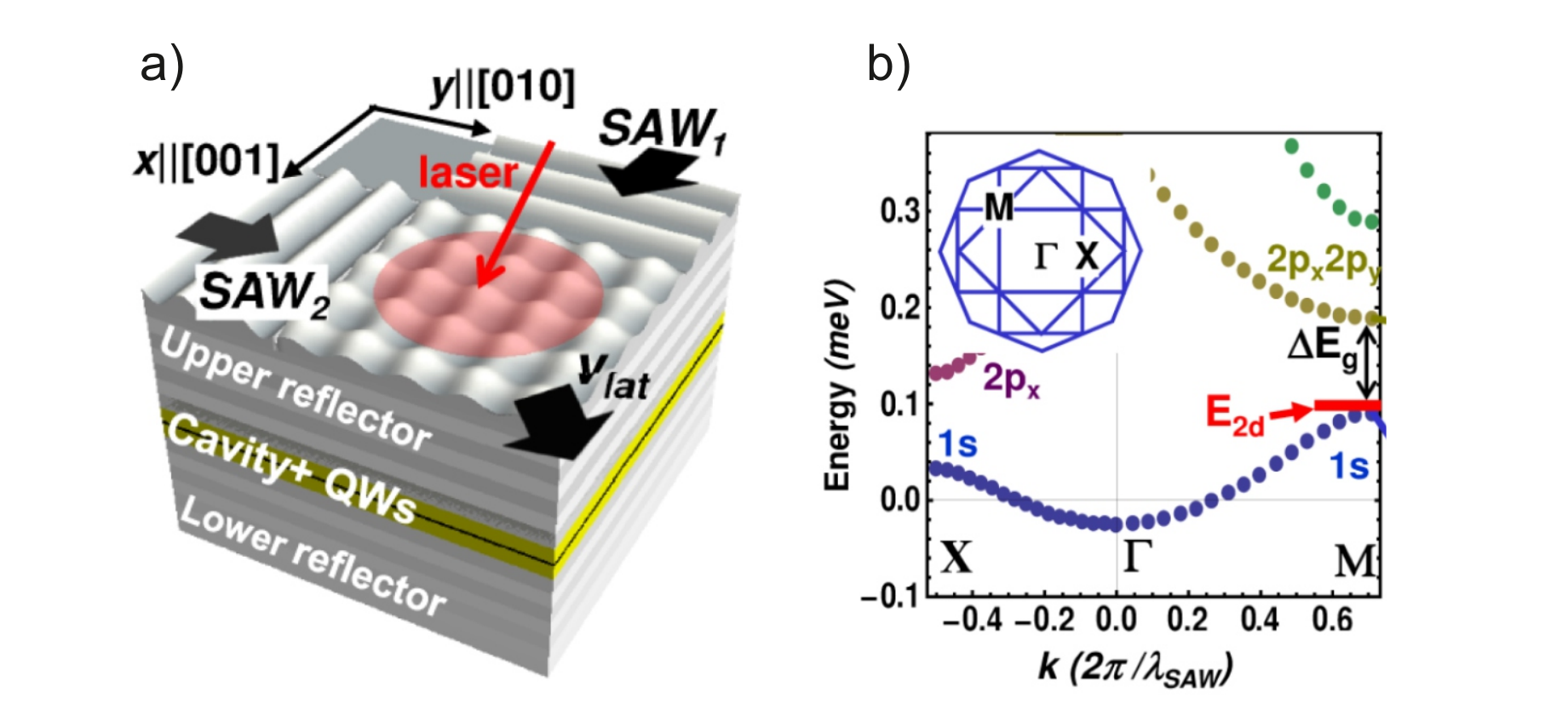}
\caption{\label{fig:Fig7} a) Schematics of two propagating SAWs, forming a square lattice potential on a polariton sample. b) Calculated bandstructure of the lower polaritons arising in such a periodic potential. The figure is reproduced from \cite{cerda-PRL13}. Copyright (2013) by The American Physical Society.
}
\end{figure}

A somewhat related technique, which also exploits a modification of the local strain environment to manipulate the polariton potential landscape, is the application of surface acoustic waves (SAWs) to the polariton system. The field can, for instance, be electrically stimulated by an interdigitated transducer 
The sound waves of the acoustic field then propagates along the sample, as sketched in Fig. \ref{fig:Fig7} a) for a two dimensional square configuration.
Strictly speaking, and similarly to the local static strain technique, the
acoustic phonon field has a direct impact on \textit{both} the excitonic and the photonic part of the polariton:
\begin{itemize}
	\item As discussed above, the strain pulse locally affects the energy of the QW-exciton via the stress depending bandgap, or more precisely, via the deformation potential interaction energy. This effect is the stronger one of the two.
	\item Both the physical cavity length and the refractive index of the cavity material are locally modified in the presence of the acoustic field, which leads to a change of the local resonance condition and hence to optical confinement.
\end{itemize}

The depth of the excitonic confinement exceeds the photonic modulation by more than a factor of 2, yet both effects on their own are sufficiently strong to provide effective polariton localization \cite{Lima-PRL06}.
Fig.~\ref{fig:Fig7}~b) depicts the influence of the SAWs on the bandstructure of the lower polariton branch. As a result of the generation of a two-dimensional polariton superlattice, a strong modification of the polariton dispersion is evidenced from the calculations, as depicted in Fig. ~\ref{fig:Fig7}b) \cite{cerda-PRL13}. In such a potential landscape, the condensation of polaritons close to the M-point of the BZ in the lattice bandgap was observed, where the effective mass of the polaritons acquires a negative sign.

\subsubsection{Proton implantation induced intermixing}

Local interdiffusion of compound QWs with the barrier material is a promising technique to generate large excitonic trapping potentials with lateral extensions below 10 $\mu$m. Intermixing techniques have been widely studied in semiconductor laser research \cite{hofstetter1998quantum}, and offer a convenient way to tailor QW properties post-growth. However, this comes at the expense of a loss in flexibility, since intermixing naturally is an irreversible  process.

One possible scheme to locally tune a QW via rapid thermal annealing without damaging its properties can be facilitated by masking the QW-cavity wafer with a material of very low heat transmittance. Polariton traps can be created at these positions after the annealing process, as the annealing procedure leads to a local redshift of the QW emission. SrF has been identified to be a good candidate for this application \cite{hofstetter1998quantum}. Underneath the masked areas the GaAs QWs (for instance embedded in AlAs barriers) remain widely unaffected by the annealing process if moderate temperatures below 900 $^\circ C$ and reasonable annealing times are applied. In the surrounding areas, which could for instance be covered by a heat transmitting dielectric or a diffusion enhancing material, such as SiO$_2$, the QW emission energy can be blue-shifted via QW-barrier intermixing.

An alternative technique to induce a local intermixing between the QW material and its barrier employs the deliberate disordering and annealing of the semiconductor heterostructure interfaces by high energy ion implantation~\cite{li2000semiconductor}. Depending on the atomic species being implanted and the dose (ions/cm$^{2}$), a rapid thermal anneal can recover the induced damage of the ion beam to a high degree.  Protons are the natural choice of ion for this technique, as hydrogen is a common, largely inactive interstitial atom incorporated during growth, and will have a negligible effect on the post-processed material quality.

\begin{figure}[ht]
    \centering
    \includegraphics[width=8cm]{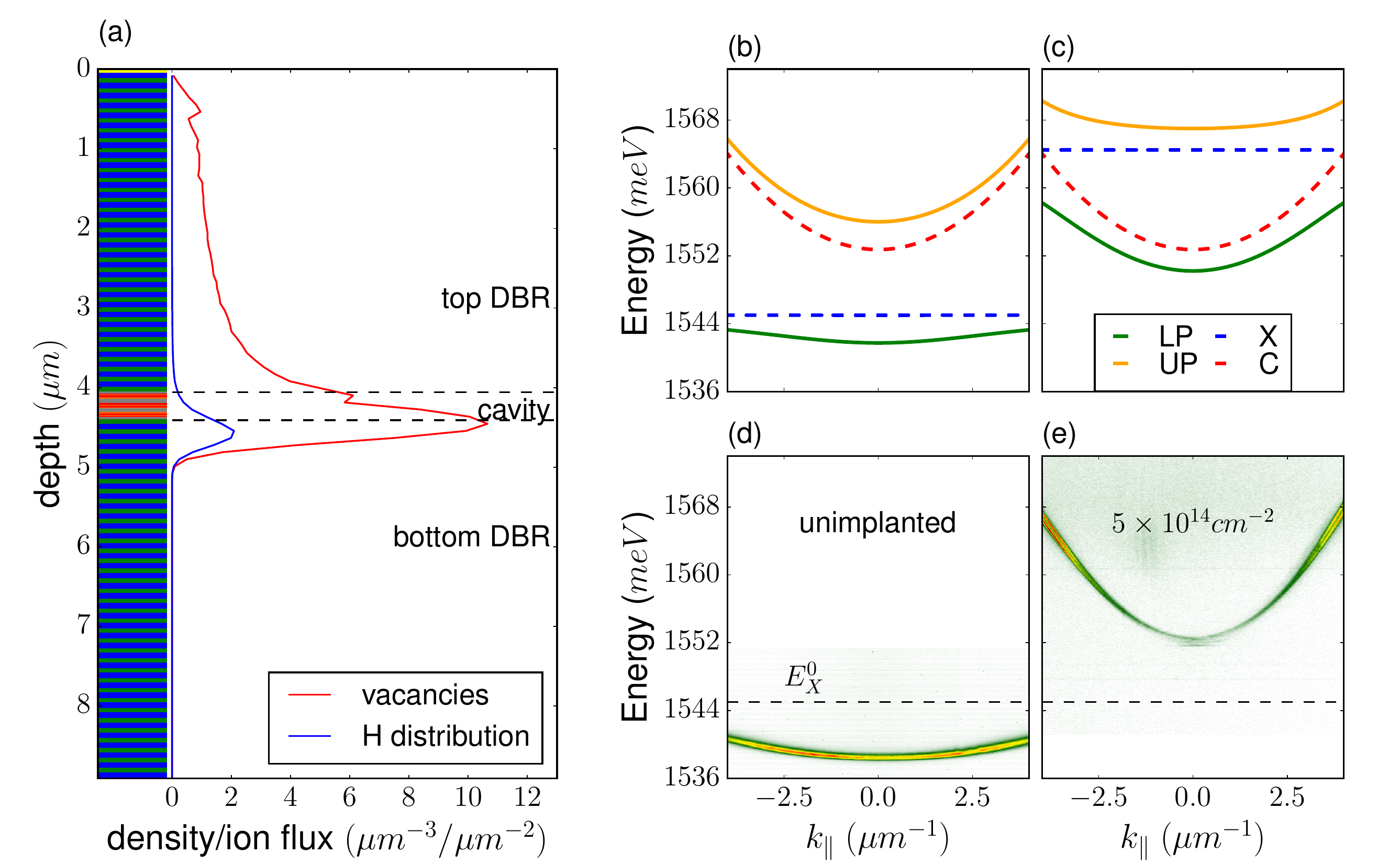}
    \caption{(a) shows the calculated ion and induced vacancy distribution at the optimal ion energy for a specific microcavity structure in order to localize the vacancies to the cavity region.  As the exciton energy $E_{X}^{0}$ is increased in energy with increasing dose, an initially blue-detuned LP (b) will also shift to higher energy (c), while its mass decreases as a result of the shift towards negative detuning.  Experimentally, an initially blue-detuned sample (d) after implantation with a dose of $5\times10^{14}cm^{-2}$ ions at $440keV$ (e), shows exactly these features.}
    \label{fig:implantation1}
\end{figure}

Despite the thickness of the top DBR layer, the ion energy can be tuned specifically such that the protons deposit the bulk of their energy in the QW-embedded cavity region, thus creating the highest density of vacancies. Figure~\ref{fig:implantation1}a) shows the simulated ion and vacancy profile in a typical GaAs microcavity structure with 24/27 mirror pairs on the top/bottom of the cavity.  For the chosen energy of 440keV, the ion and vacancy concentration is largely localized to the cavity region.

The applicability of this technique is strongly supported by previous studies~\cite{tan1996large} of proton implantation-induced intermixing on bare GaAs/AlGaAs QWs which show energy shifts in the 100's of meV range, and specifically, that at smaller doses, the luminescence intensity and linewidth can be recovered close to original values but still with 10s of meV shifts, sufficient for a tight and deep trapping of exciton-polaritons in microcavities.

The shift to the lower polariton (LP) state, will strongly depend on the initial cavity-exciton detuning.  Given a blue-detuned LP state (fig.~\ref{fig:implantation1}b), as nominal exciton energy $E_{X}^{0}$ is shifted to the blue (and correspondingly the LP energy) with increasing dose, the lower polariton detuning will shift to the red (fig.~\ref{fig:implantation1}c).  As far red-detuned polaritons are quite insensitive to the exciton energy, the LP mode will only shift to a maximum energy roughly equivalent to its amount of positive detuning.  Provided the Rabi energy is also sufficiently large, implanting a low dose of protons into a polariton microcavity, with subsequent thermal annealing, will create a strongly confining potential for a LP state with blue detuning.

Initial experimental studies
of the technique on planar samples can reproduce the above described features.  From an originally blue-detuned sample piece (fig.~\ref{fig:implantation1}d), when implanted with a moderate dose, yields a polariton mode shifted by $\Delta E_{LP}\sim$15meV, and showing a large reduction in the effective mass consistent with a shift of the polariton detuning into the red detuning region (fig.~\ref{fig:implantation1}e).

\subsubsection{Electrostatic Traps}

\begin{figure}[ht]
\centering
\includegraphics[width=8cm]{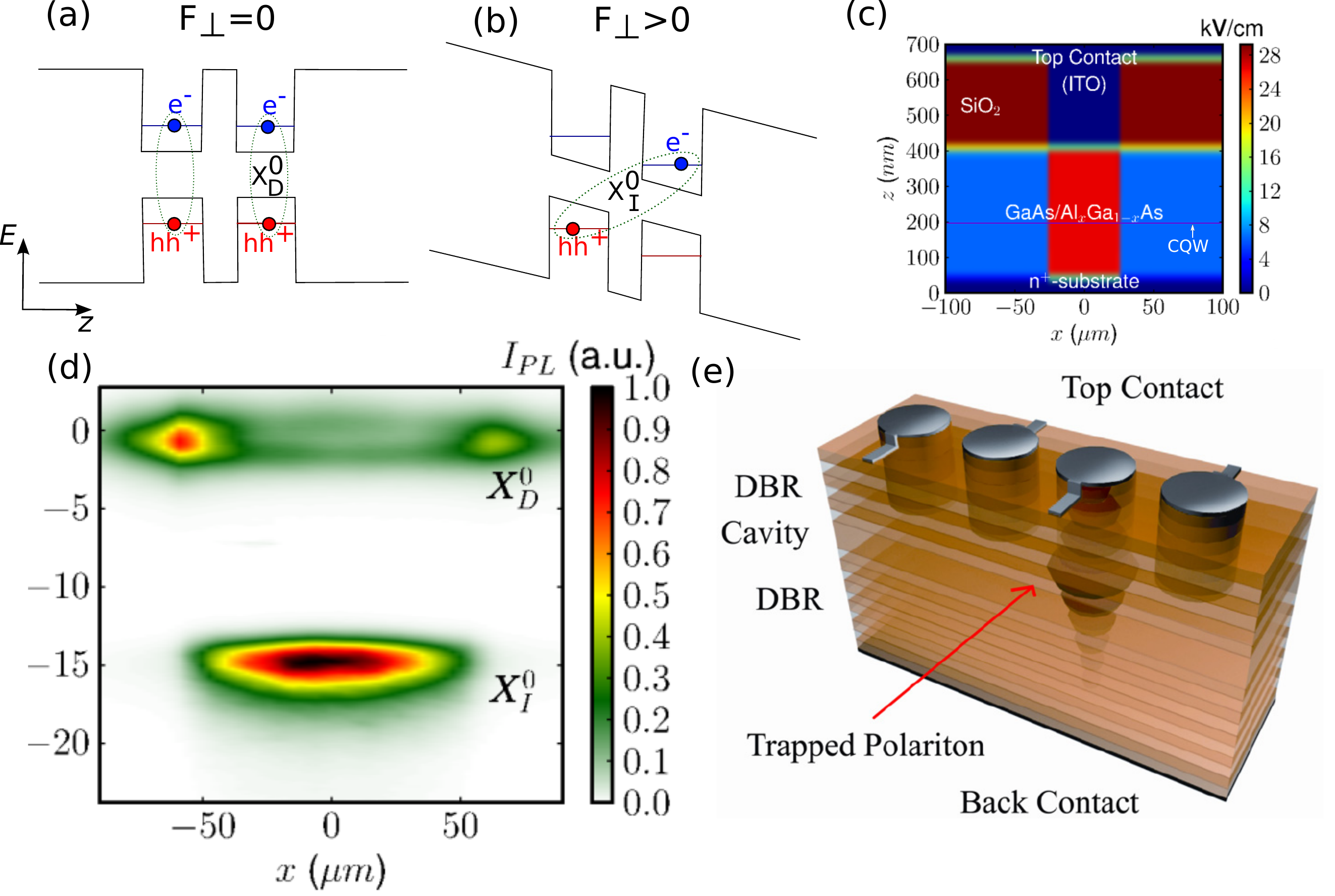}
\caption{\label{fig:Fig9} Electrostatic trapping of excitons and exciton-polaritons: a) and b) The exciton radiative transition without and with an applied voltage, the latter with long-lifetime indirect excitons $X_{I}$. c) A cross-section of a trap design showing a modulation of the electric field strength felt by the excitons due to shielding by an SiO$_{2}$ layer, and d) the experimental realization of indirect exciton trapping. These figures are reproduced from \cite{fraser2011selective}.  Copyright (2011) by The American Physical Society. e) Proposed trap for microcavity exciton-polaritons: The microcavity is locally capped by a semitransparent contact which simultaneously acts as a gate. Once a small bias is applied, the polariton energy is lowered underneath the trap, and an attractive potential is created. The figure is reproduced with permission from \cite{na2010massive}}

\end{figure}

By applying electric fields to semiconductor quantum wells, the energy of the exciton can be tuned via the quantum confined Stark effect (QCSE). When locally applied to a sample, this effect can also be exploited to build flexible and elegant exciton traps. A major advantage of this technique certainly lies in the ability to electrically manipulate the trapping potential.
For weak electric fields $F$, the shift of the exciton energy due to the QCSE can be expressed as:
\begin{equation}
\Delta E= \beta F^2,
\end {equation}
where $\beta$ is the polarizabiliy of the quantum well excitons. Note that, for indirect excitons in coupled QWs (these are excitons where the electron is confined in one QW, and the hole in the other QW \cite{zrenner1992indirect}, see Fig.~\ref{fig:Fig9}~a,b) or for excitons in thick QWs, for high fields  the energy shift is linear with the applied electric field $\Delta E= -e d_{eff} F$, where $d_{eff}$ denotes the separation between the QWs. This trapping technique has been successfully exploited to localize (mostly indirect) excitons to study cold exciton gases in the presence of confinement: A cross section of a design for such a trap (reproduced from \cite{fraser2011selective}) is shown in Fig.~\ref{fig:Fig9}~c).The electric field strength is laterally modulated due to shielding by an SiO$_{2}$, which leads to the distinct formation of an electrostatic trap of QW excitons as can be seen by the deeply trapped indirect excitons in Fig.~\ref{fig:Fig9}~d) . This technology has been applied in similar forms for the creation of exciton traps with complex geometries \cite{Gartner-PRB07}, yielding the possibility to even promote bosonic condensation phenomena \cite{Butov-Nature02}.

A possible scheme to fabricate electrostatic polariton traps is shown in Fig~\ref{fig:Fig9}~e): Semitransparent contacts are fabricated on top of a doped microcavity. Since the QWs are located in the intrinsic region, moderate applied voltages of a few hundred mV should suffice to shift the QW exciton emission up to several meV. It is interesting to note that, while the QCSE has been studied for exciton-polaritons in the linear regime \cite{Fisher-1996, Gessler-APL14, Brodbeck-OE13}, certain peculiarities related to phonon enhanced tunneling effects between nearby QWs and carrier screening were observed recently \cite{Tsotsis-PRA14, Brodbeck-APL15}. Under specific circumstances, in particular if diffusion of the reservoir is restricted, these effects indeed lead to a blueshift of the polariton resonance in the presence of the electric field, which may present a complication for the realization of electrostatic polariton traps.

An intrinsic limit to the tuning range of the exciton emission is carrier tunneling out of the finite QW barriers. Furthermore, in order to laterally trap excitons, the QCSE has to be locally applied via a finite-size gate. As a result, lateral electric fields will evolve in the QW \cite{Hammack-JAP06}, which can also lead to a dissociation of the excitons.
Despite the fact that there still is no conclusive demonstration of an electrostatic polariton Stark trap to date, the development of this technique would be highly advantageous, in particular because of its flexibility. A combination of electrostatic trapping or tuning with other confinement techniques described later in this article could yield the possibility to engineer reconfigurable polariton landscapes with unprecedented properties.

\subsubsection{Confinement provided by the exciton reservoir}
\label{optical-traps}

\begin{figure}[ht]
\centering
\includegraphics[width=8cm]{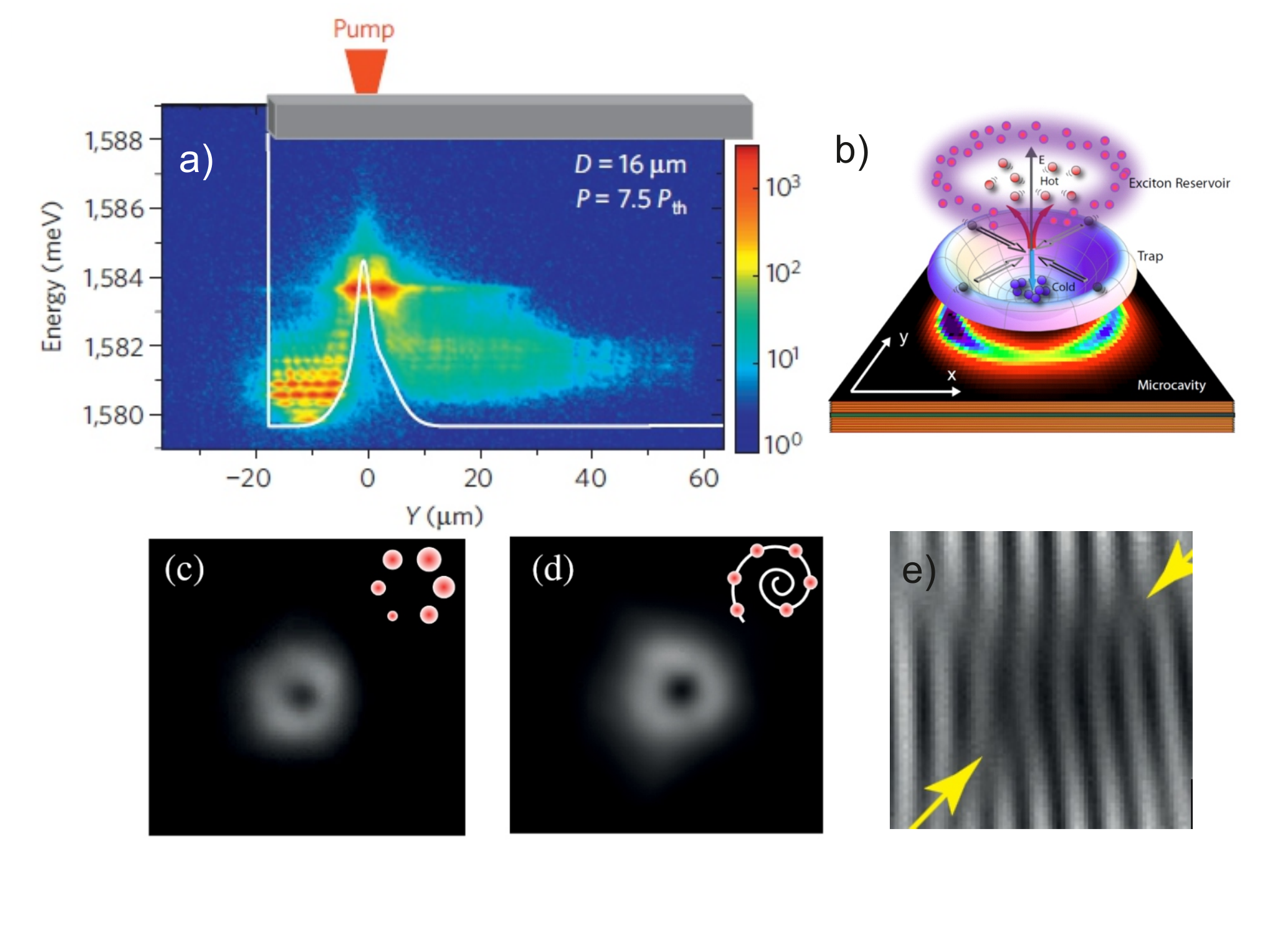}
\caption{\label{fig:flow} All-optical trapping of polaritons with the pump laser. a) The polaritons are already confined in a one-dimensional microwire. The excitation laser spot provides the remaining boundary, and a zero-dimensional polariton condensate is formed close to the edge of the wire. The figure is reproduced from \cite{Wertz-NatPhys10} b) With a ring-shaped excitation laser spot, it is possible to create and confine a condensate in the center. The figure is reproduced from \cite{Askitopoulos-PRB13}.  Copyright (2013) by The American Physical Society. c-d) Method to create chiral structures by tailoring the exciton reservoir with a structured pump beam, by misalignement (c) and deliberatly designing the pump structure (d). Both techniques create a charge on vortex via non-resonant pumping. e) Interference image with a characteristic fork pattern, indicating the presence of  a vortex. c-e) are reproduced from \cite{Dall-PRL2014}.  Copyright (2014) by The American Physical Society.}
\end{figure}

The carrier and exciton reservoir, which is induced by the excitation laser, provides a natural way to create potentials for the exciton polaritons. This approach provides a very elegant and useful technique to confine, manipulate and steer polaritons by exploiting their strong Coulomb interactions. An initial demonstration of this effect was discussed in a one-dimensional configuration \cite{Wertz-NatPhys10}. Here the polaritons were already partly confined in a deeply etched wire cavity along the wire. The reservoir created by the injection laser provided an additional potential barrier, resulting in polariton confinement in all spatial directions close to the edge of the wire (Fig.\ref{fig:flow}a). Conveniently, the size of the trap can be modified by changing the location of the pump laser spot on the wire cavity. In a similar way, polaritons can be confined all-optically by choosing smart geometries of the excitation laser beam. By using spatial light modulators or shaping the optical excitation by means of an amplitude mask, ring shaped confinements \cite{Askitopoulos-PRB13} (see Fig. \ref{fig:flow}b), confinements created by multiple pump spots \cite{Cristofolini-PRL13, Tosi-NatPhys12} and controlled flow patterns in more complex landscapes were generated \cite{Schmutzler-arxiv14}.
The technique to structure the (non-resonant) pump profile has been extented in \cite{Dall-PRL2014} to deliberately produce polariton condensates with defined chirality. There, the pump beam has been structured via a metal mask with six pin-holes, which either were misaligned with respect to the laser beam, or structured in a chiral arrangement. Both techniques were proven useful to excite trapped vortices in the center of the six excitation spots, as confirmed by real space interferometry (see Fig.~\ref{fig:flow}c-e).

The technique of controlling the environment via shaping the pump profile has proven extremely useful for the manipulation of the polaritons' properties due to its flexibility, however the depth of the provided confinement (or the potential height) is limited by the strength of the polariton-exciton repulsive interaction, which is of the order of 1 meV.

\subsubsection{Low-dimensional active material}

In order to complete this section, we will briefly review some examples of creating microcavity polaritons by directly using low-dimensional gain material instead of two-dimensional quantum wells subject to lateral trapping. In this discussion, we will limit ourselves to single semiconductor quantum dots on GaAs.
A system composed of a single QD and a microcavity can be described by the single-particle Jaynes-Cummings Hamiltonian \cite{Michler-2}, giving rise to an energy ladder structure which adds strong non-linearities on the single photon level. For this reason, QD-polaritons are usually studied in the framework of isolated quantum systems, similar to atoms in high finesse microcavities \cite{miller2005trapped}. Naturally, the possible application of QD-polaritons fundamentally differ from the QW-case.

Successful realizations of QD-polariton systems were first reported in \cite{Reithmaier-Nature04, Yoshie-Nature04, Peter-PRL05} for various photonic resonator geometries. The first successful demonstration of true quantum effects in such systems (i.e. sub-poissonian emission statistics) in  fundamental contrast to quantum well microcavity polaritons, was reported by \cite{Press-PRL07} and \cite{Hennessy-Nature07}. Signatures of the photon blockade (or QD-polariton blockade), which is a fundamental effect which occurs due to the single photon non-linearity of the Jaynes-Cummings ladder system (see \cite{kasprzak2010up}), were observed in \cite{Faraon-NatPhys08}. The electrical injection \cite{Kistner-APL10}, electro-optical manipulation \cite{Kistner-OE08} and nonlinear emission \cite{Nomura-Naturephysics10} have been reported.

While the strong exciton localization in the quantum dot structures has many advantages due to the inherently very large non-linearities on the single- to few photon or exciton scale (leading to single photon emission and the polariton blockade effect), its main drawback lies in the poorly controlled  nature of the fabrication process. It is still very challenging to gain control over the properties of single quantum emitters to such an extend, that they can be e.g. arranged in arrays \cite{Schneider-Nanotechnology09, Schmidt-Buch} with designed spectral properties \cite{Mereni-APL09}. In this respect, a QW-polariton system, if it could be designed in such a way that nonlinearities are strongly enhanced, would be highly preferable.

\subsection{Trapping of polaritons via their photonic part}

\begin{figure}[ht]
\centering
\includegraphics[width=8cm]{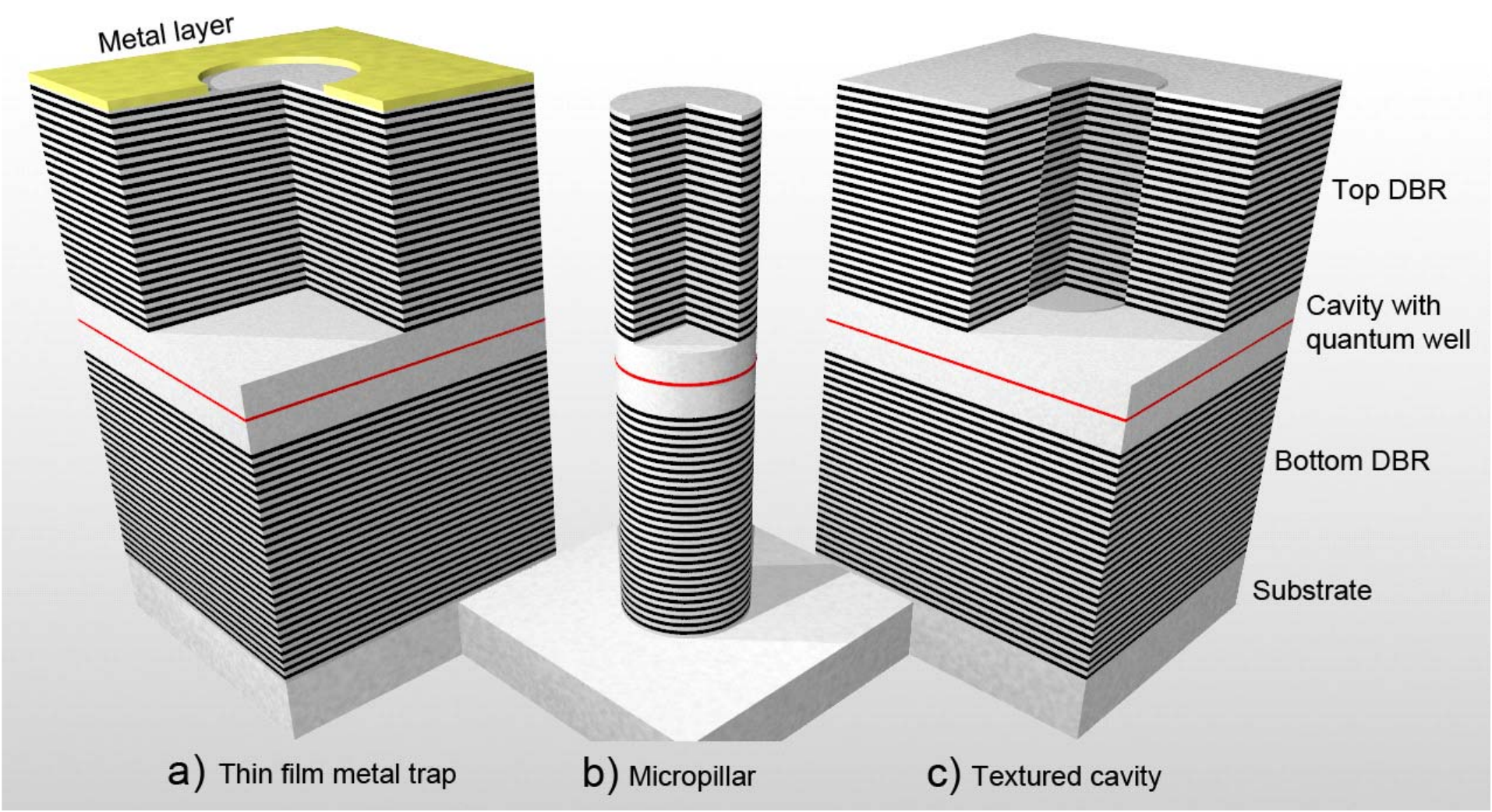}
\caption{\label{Abb:Fig7} Schematics of commonly applied techniques to spatially confine exciton-polaritons via the photonic part: a) depositing metal films on the top DBR layer, b) etching micropillars and c) defining shallow mesas in the cavity.
}
\end{figure}

Complementary to the discussion on excitonic polariton trapping, in the following we will consider techniques to confine polaritons via their photonic part. Due to the relatively large extent of the photonic wavefunction, quantization and finite size effects can be observed for relatively large structures on the order of the polariton wavelength. This gives rise to 'quasi zero dimensional' exciton polartions in structures with lateral dimensions on the order of 1-10 $\mu$m. Three of the most frequently exploited techniques, namely etching micropillars, defining shallow mesas and depositing metal masks on top of the two dimensional microcavity are sketched in fig.~\ref{Abb:Fig7}, and will be discussed in more detail in the next section.

\subsubsection{Natural photonic disorder traps}
\label{natural}

In many microcavities, polariton localization can inherently occur in so-called photonic disorder potentials. These potentials can arise, e.g., due to natural local elongations of the cavity at the position of crystal defects and provide an optical confinement. It has also been shown, that the condensation of polaritons in a planar two-dimensional cavity is likely to occur in such natural potentials \cite{Kulakovskii-JETP10}. An accurate characterization of the formation of polaritonic states in natural traps was performed in \cite{Zajac-PRB12}. Interestingly enough, these defect-induced traps are typically of a Gaussian shape, which is predicted to promote confined Q-factors strongly exceeding those of their mesa (vertical solid wall)  counterparts for comparable mode volumes \cite{Ding-PRB13}. This makes such natural crystal defect traps highly appealing for the demonstration of large polariton non-linearities and even photon blockade effects \cite{Verger2006}. However, the scalability of this approach is strongly limited due to the random nature of these defects. In a later subsection of this article, we will discuss in detail a technology enabling to implement similar structures in a microcavity in a fully controlled manner.

\subsubsection{The metal mask technique}

\begin{figure}[ht]
\centering
\includegraphics[width=8cm]{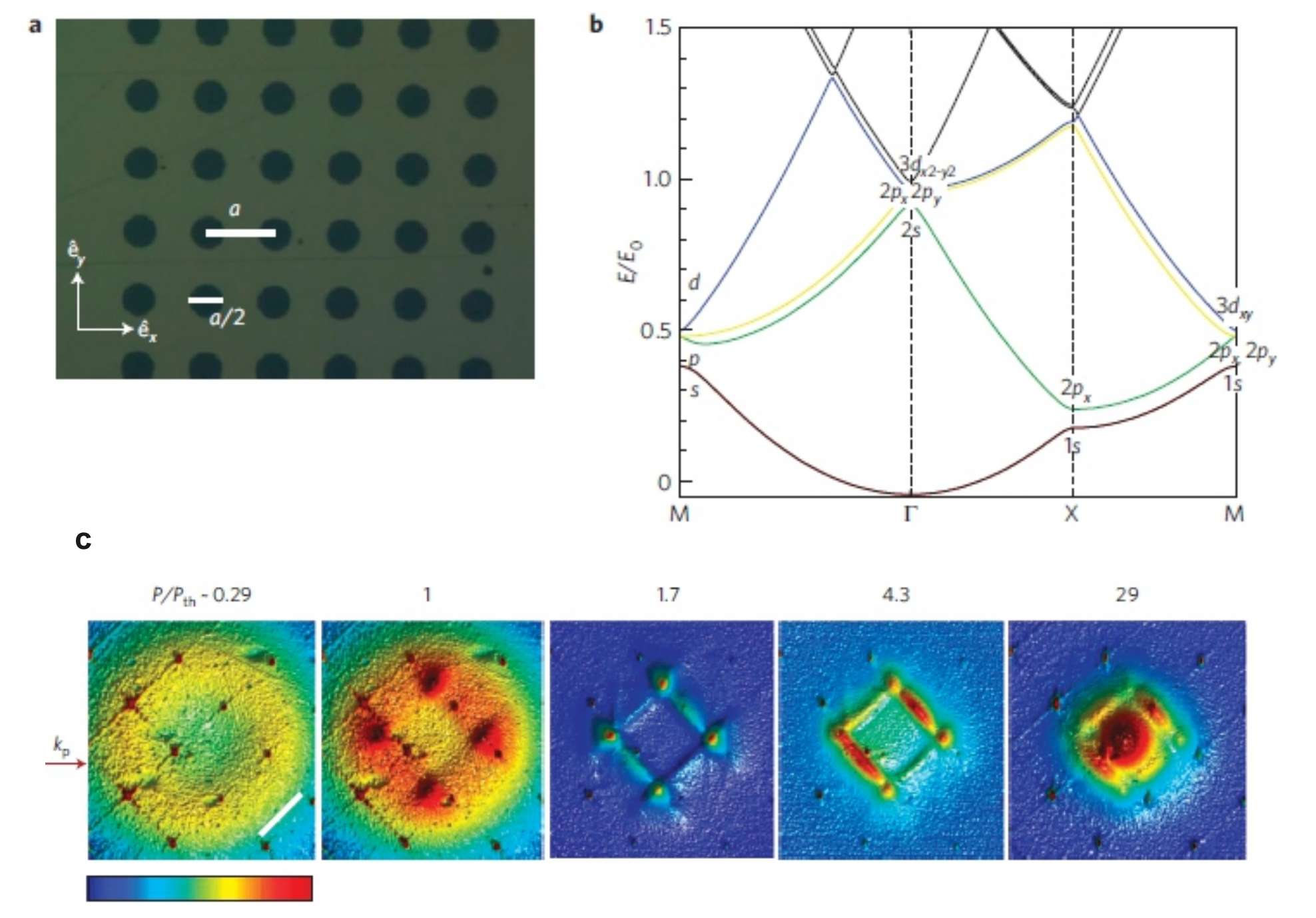}
\caption{\label{Abb:Fig15} Technology to confine polaritons by the deposition of thin metal layers on a microcavity: a) Scanning electron microscope (SEM) image of a metal layer with micron sized circular holes forming a square lattice arrangement. b) Calculated energy-momentum dispersion relation for the given geometry. c) Demonstration of the condensation of polaritons close to the high-symmetry points in the BZ, starting with a condensate at the characteristic M-point of the BZ. Reprinted with permission from \cite{Kim-NatPhys11}.
}
\end{figure}

A comparatively simple, yet very efficient method to create polariton confinement in a grown microcavity structure is the deposition of metal films on the sample surface. The metal layer changes the boundary conditions of the electromagnetic field with respect to a semiconductor-air interface and creates an optical node. As a result, a modest shift in the energy of the optical resonance can be observed (on the order of 0.1-1 meV), which leads to an effective photonic confinement. The beauty of this approach certainly lies in its simplicity, and many pioneering experiments with polaritons in periodic potential landscapes were initially carried out in such samples. Examples include the first demonstration of a polariton condensate at higher-order Bloch bands in one-dimensional arrays \cite{Lai-Nature07} and the formation of polariton condensates in d- and f-states \cite{Kim-NatPhys11}, as well as the condensation of polaritons close to Dirac points in triangular configurations \cite{Kim-NJP2013}. A compilation of results obtained in the square lattice geometry is shown in Fig. \ref{Abb:Fig15}.

An important extension of this approach is related to the excitation of the so-called Tamm plasmon (TP) states at semiconductor-metal interfaces. Localized TPs can evolve at the crystal surface at the interface between a periodic dielectric structure and a metal layer \cite{Kaliteevski-PRB07}, and can be directly optically excited. The electric field distribution of such a TP decays into the periodic DBR structure, with a significant field enhancement close to the semiconductor-metal interface. It is rather straightforward to couple TP modes to matter excitations, and the formation of Tamm plasmon exciton-polaritons has been reported \cite{Gessler-APL14, Symonds-APL09}. In stark contrast to the deposition of the metal on top of a DBR, the lateral confinement provided by the Tamm plasmon can be significant \cite{Gazzano-PRL11}.

For both methods, the Tamm plasmon approach or the metal mask deposition (which can also be considered as evanescent coupling of a Tamm state to a cavity photon), one has to compromise between the depth of the polariton confinement and the cavity Q-factor. While the Q-factor increases with the amount of mirror pairs deposited onto the top DBR segment, the confinement depth decreases at the same time. This issue renders this approach incapable of providing spatial confinement for high-quality DBR based microcavities. For instance, for a sample with 16/20 AlGaAs/AlAs mirror pairs with a Q-factor on the order of 2000, a confinement depth of about 200 $\mu$eV was experimentally extracted \cite{kim2008gaas}. This number would be reduced to a value between 10-50 $\mu$eV for a sample capped by more than 25 mirror pairs, as can be determined by a standard Transfer Matrix calculation.

\subsubsection{Etched Micropillar cavities}

\begin{figure}[ht]
\centering
\includegraphics[width=8cm]{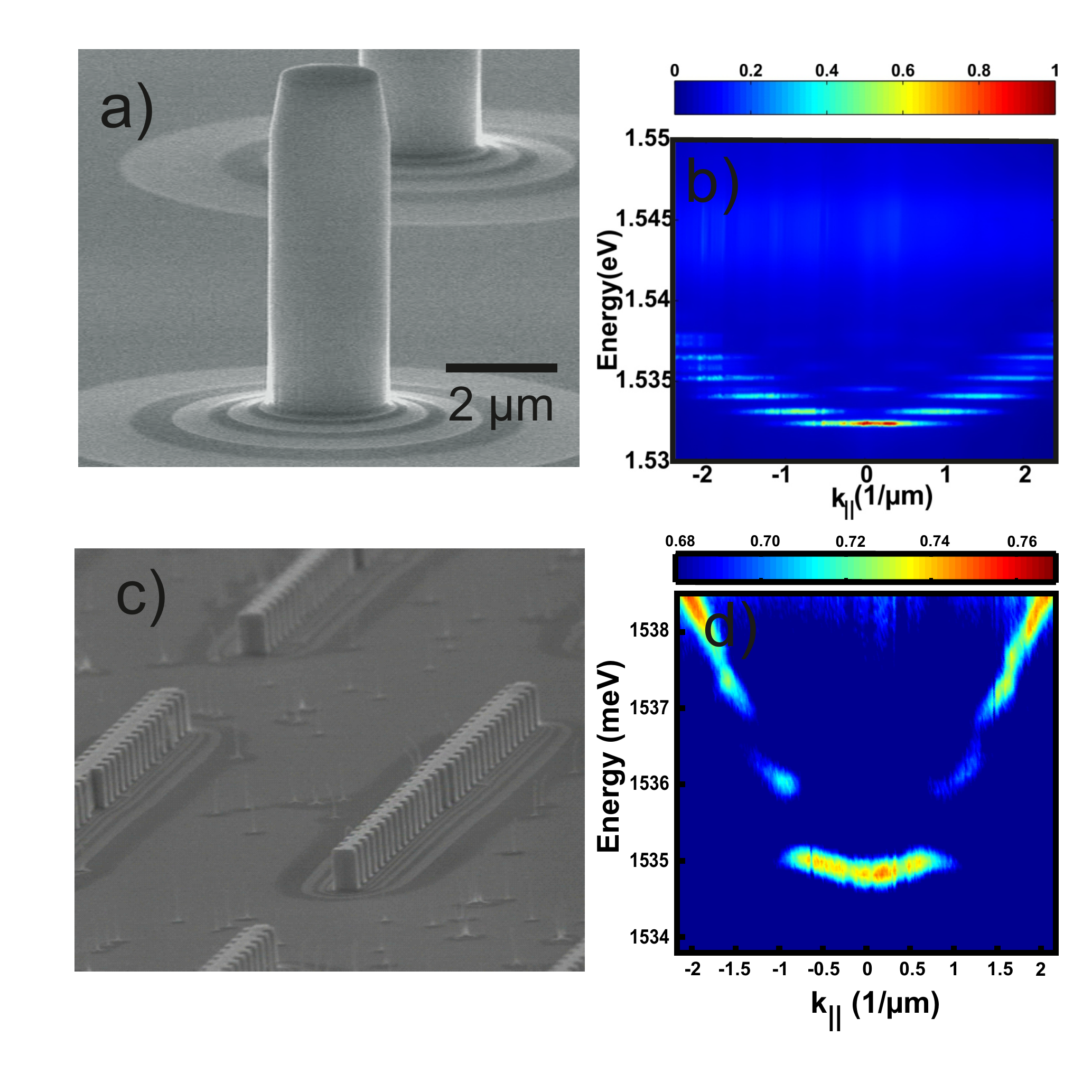}
\caption{\label{Abb:Fig10} a) SEM image of a micropillar cavity, etched into a DBR based structure. b) Photoluminescence spectrum of the optical modes in a micropillar device. c) SEM image of a 1D arrangement of rectangular micropillars, resembling a modulated microwire. d) The energy-momentum dispersion of the modulated wire features a band gap spectrum with distinct gaps at the edges of the BZ.
}
\end{figure}

Deep etching of micro pillars in the sample is perhaps the most straightforward approach to lateral confinement of a photon in a DBR microcavity. Additionally, the diffusion and propagation of excitons is also limited to the pillar size, which has some important implications on the performance of polariton lasers in the pillar geometry \cite{kochereshko2013phase}. In this geometry, the optical field is laterally confined by the semiconductor-air interface. This approach has been widely exploited to fabricate vertically emitting lasers and efficient single photon sources \cite{Moreau-APL01, Santori-Nature02, Heindel-APL10}. In order to realize such structures, typically optical lithography or electron beam lithography is employed to define an arbitrary shape of the pillar in a photo- or electron beam sensitive resist. This resist can be either directly used as an etch mask, or the latter can be evaporated after the lithography step. Plasma etching has been proven to facilitate the realization of nearly perfect posts with vertical pillar sidewalls in the GaAs system. A typical example of such a micropillar with a diameter of 2 $\mu$m is shown in Fig. \ref{Abb:Fig10}a). Note that careful optimization of the etching procedure (e.g. reactive ion etching with $Cl_2$/Ar plasma) can lead to almost perfectly vertical and smooth sidewalls, which is crucial when scattering and diffraction losses should be kept minimal \cite{Schneider-arxiv2015}. With this, it is possible to  realize DBR-based zero dimensional microcavities with Q-factors exceeding 150000 \cite{Schneider-arxiv2015}. The finite physical size of this micropillar results in a strong optical confinement for the microcavity photon as a result of the huge difference between the refractive index in the structure (n$\approx 3.5$) and its surrounding (n$\approx 1$). Hence, the optical mode spectrum splits into a set of characteristic waveguide modes. A characteristic photoluminescence spectrum from such a micropillar cavity is shown in Fig. \ref{Abb:Fig10}b). The parabolic cavity resonance splits into a set of discrete, confined photonic modes, arising from the circular waveguide geometry (see, e.g. \cite{Silva-OE08, Ulrich-PRL07, Reitzenstein-JPD10}).

Despite the brute force approach to create lateral optical confinement, and the associated degradation of the QW properties via surface related effects, the first demonstration of strong coupling effects in such micropillars was reported by Gutbrod et al. \cite{Gutbrod-PRB98}, where the characteristic anticrossing between optical and electronic resonance was fully mapped out via temperature tuning. In this work, the effects of light-matter coupling were amplified by an applied magnetic field, which enhanced the oscillator strength of the integrated QWs and lead to an increase of the light-matter coupling strength.  Other benchmark experiments include the first demonstration of polariton condensation in a micropillar cavitiy in the II/VI system \cite{Obert-APL04} and the GaAs system \cite{Bajoni-PRL08}. \\

A particular advantage of defining polariton potentials via lithography and etching lies in the  simplicity of the approach. It is comparably easy to create almost arbitrary potential profiles and to study the behavior of the quantum gas in such tailored environments. In this spirit, polariton condensates in one dimensional microwires have been generated and their propagation and coherence properties were investigated \cite{Wertz-NatPhys10, Schmutzler-PRB14, Fischer-PRL14}. Furthermore, the condensation of polaritons in microwires with a modulated sidewall (in other words, a one-dimansional array of coupled rectangular micropillars (see Fig.\ref{Abb:Fig10}c) has been reported \cite{Tanese-NatComs13}. The overlap between adjacent pillars leads to photonic coupling, resulting in photonic bands with well pronounced gaps at the edge of the BZ (Fig.\ref{Abb:Fig10}c), which has been investigated already in the early works by Dasbach et al. \cite{dasbach2003spatial}. As we will detail later, the peculiar band structure of such an array of coupled micropillars featuring full bandgaps as well as areas with negative effective masses give rise to interesting condensation phenomena, such as the formation of gap solitons \cite{Tanese-NatComs13}.   \\

\begin{figure}[ht]
\centering
\includegraphics[width=8cm]{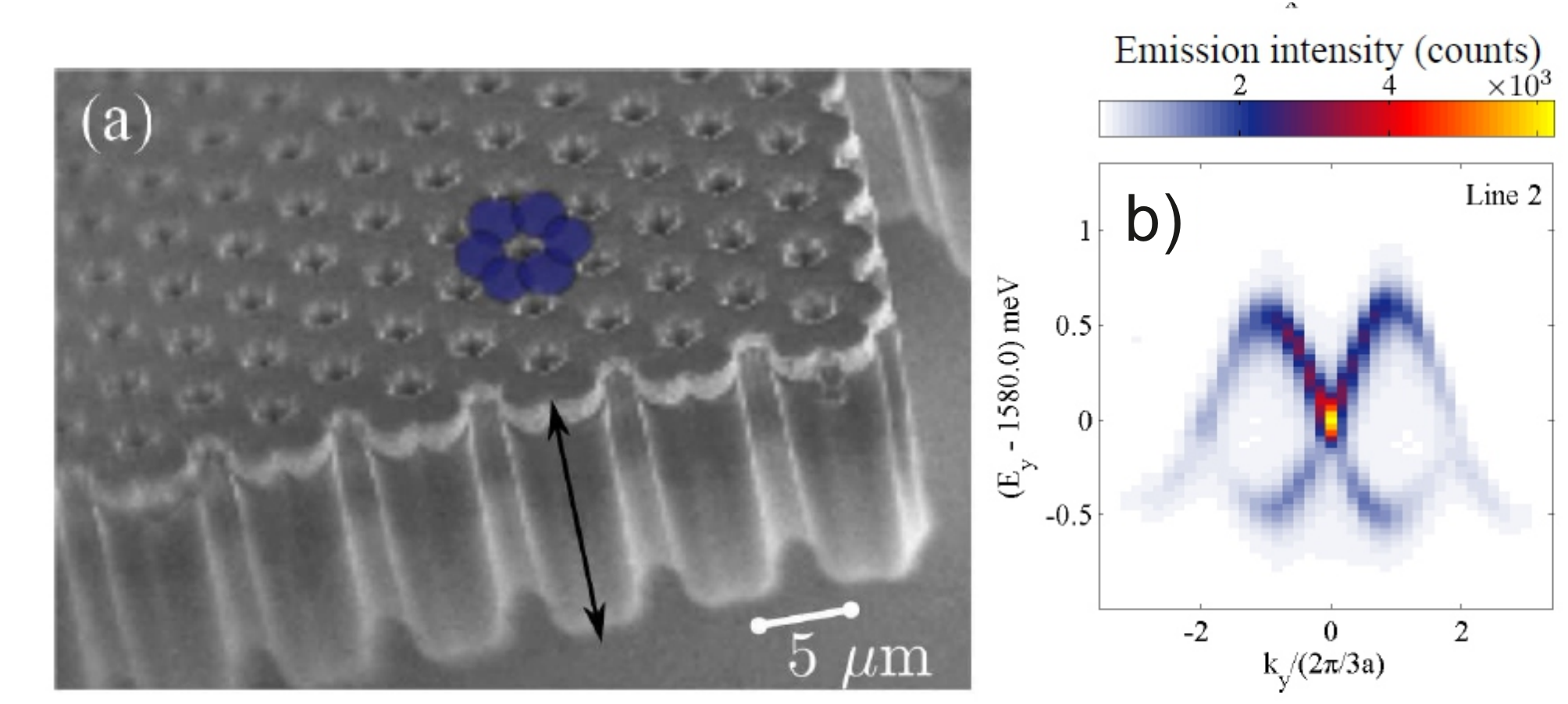}
\caption{\label{Abb:Fig11} a) SEM image of an array of micropillars aligned in a honeycomb lattice configuration. b) The potential landscape gives rise to Dirac cones with a linear Energy-momentum dispersion relation. The condensation of polaritons near such a Dirac point is discussed in \cite{Jacqmin2014}. Copyright (2014) by The American Physical Society.
}
\end{figure}

Owing to the high flexibility and maturity of the electron beam lithography and etching technologies, two-dimensional arrays of overlapping micropillars can also be realized rather straight-forwardly. Initial measurements of the photonic band structure in two dimensional micropillar lattices were reported by Bayer et al. \cite{Bayer-PRL99}. An example of the implementation of a polariton condensate in a hexagonal potential environment is shown in Fig. \ref{Abb:Fig11}a). The structure features an optical potential landscape resembling the graphene-type lattice (so-called honeycomb structure), which is also reflected in the optical properties of the emitted light from the bosonic system. The most striking property of this lattice configuration is the appearence of the so-called Dirac points at specific locations in the BZ. A Dirac point is characterized by its linear dispersion of mass-less particles. The condensation of polaritons in the vicinity of a Dirac point is demonstrated in Fig. \ref{Abb:Fig11}b).

 Furthermore, the maturity of this approach has already led to the first generation of polaritonic logic devices designed for integrated photonic experiments. These devices take advantage of the possiblity to direct the flow of polaritons or condensates along etched channel structures over macroscopic distances, and to manipulate them via optical or electric fields, resulting in the successful implementation of switches, interferometers and tunneling diodes \cite{gao-PRB12, Sturm-NatComs14, nguyen-PRL13}.

The ease of the micropillar approach comes along with a significant  disadvantage. In order to provide strong mode confinement, the structures are commonly etched through the active region (the QWs). As a consequence, nonradiative recombination of the excitons at the surface can occur, and furthermore etching through the QWs is accompanied by a degradation of the emission. This effect can be  pronounced, and is visible for instance in the  spectrum of the etched micropillar in Fig.~\ref{Abb:Fig10}b), where emission from a strong uncoupled background resulting from etching through the active region can be seen. Therefore, less destructive methods to incorporate deep and flexible polariton confinements into polariton landscapes are highly desirable.

\subsubsection{Photonic crystals and hybrid approaches}

\begin{figure}[ht]
\centering
\includegraphics[width=8cm]{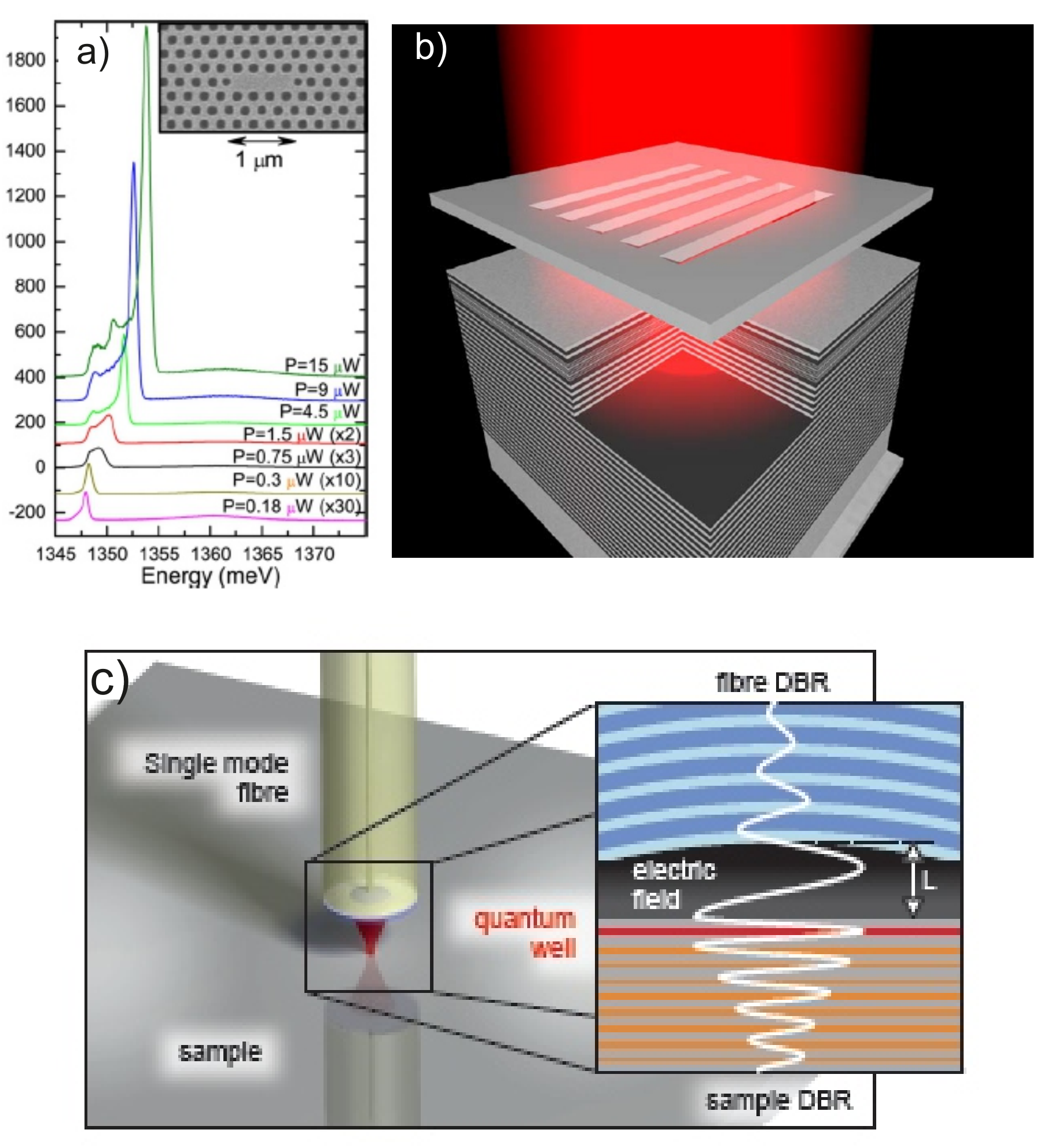}
\caption{\label{Abb:Fig12} a) Polariton laser based on a GaInP L3 photonic crystal nanocavity (Figure reprinted with permission from Azzini et al. \cite{Azzini-APL11}). b) Sketch of a hybrid photonic crystal-DBR polariton laser. c) Hybrid fiber cavity polariton laser. The figure is reproduced from \cite{besga2015polariton}. Copyright (2015) by The American Physical Society.
}
\end{figure}

Similarly to the fabrication of micropillar cavities to confine polaritons, photonic crystal nanocavities can also provide a tight optical confinement with the smallest mode volumes. This technique is particularly popular for the fabrication of microlasers with the smallest footprints and coupled systems with quantum dot emitters, due to the small mode volumes and the resulting large light-matter coupling strength \cite{Yoshie-Nature04, Hennessy-Nature07}. To date, there are only few demonstrations of polariton lasers in conventional photonic crystal nanocavities \cite{Azzini-APL11} (Fig. \ref{Abb:Fig12}a). A (hybrid) photonic crystal approach which fully circumvents the etching of the active medium has been introduced by Zhang et al. \cite{Zhang-Light14, Fischer-APL14}. Similarly to the methods used to build flexible vertically emitting microlasers \cite{Huang-NatPhot07}, only the upper DBR mirror is replaced by a highly reflecting sub-wavelength high contrast grating (HCG), which is a broadband crystal mirror. A sketch of the resulting structure, which has been proven to promote polariton lasing features \cite{Zhang-Light14} is shown in Fig.\ref{Abb:Fig12}b). This work has been successfully extended by coupling of such polaritonic boxes to photonic molecules and one-dimensional superlattices \cite{Zhang-APL15}, which demonstrates the versatility and potential of the approach.

In the HCG-cavity, strong coupling conditions can only be maintained underneath the finite sized grating. This finite size effect results in an effective in-plane confinement of polaritons, restricted to the dimension of the HCG. Another approach to define zero dimensional polaritons by reducing the lateral area was introduced in \cite{besga2015polariton}. The authors designed a single mode fiber with an integrated Bragg grating which they used to replace the top DBR. While the curvature of the fiber tip defines the mode volume in such a cavity, the resonance frequency of the resulting hybrid device can be almost arbitrarily varied by varying the distance between the fiber tip and the sample surface. This elegant techniques promises ultra-high Q-factors, low mode volumes and, consequently, is a prime candidate for investigating non-linearities on the single polariton scale. A conceptually similar concept was introduced in \cite{dufferwiel2014strong}, where the authors demonstrated strong coupling conditions in a so-called open cavity approach. Here, one DBR is attached to piezo actuators which allow to tune the physical cavity length, while a tight optical confinement is introduced by locally shaping the curvature of the mirror.

\subsubsection{Etch-and-overgrowth technique}
\label{etch-and-overgrowth}

As indicated in section~\ref{natural}, a shallow modulation of the cavity length, e.g. via a crystal defect initiated by a droplet, is sufficient to create a well defined, deep lateral photonic confinement \cite{Zajac-PRB12}. A straightforward lithographic implementation of this concept leads to creation of polariton mesa traps. These traps have been first reported by El Daif et al.\cite{Daif2006} and where subsequently theoretically analyzed with respect to the feasibility to observe polariton BEC states in lattice configurations \cite{Boiko2008}. Although the realization of such mesa traps is certainly more challenging compared to other approaches, including etching of regular micropillars or deposition of  metal masks, the approach is very appealing due to the following significant advantages: (i) the confinement depth can be tuneable in a wide range, mainly by adjusting the height of the defect and the light-matter detuning; (ii) inter-site coupling between neighbouring traps is readily controllable; (iii) surface recombination effects from etching through the active medium are fully avoided.  These traps are commonly realized by texturing of the cavity layer of a Fabry-Perot-microcavity structure into elongated (trap) and regular (planar) regions. As the cavity resonance condition is fulfilled for photons with a longer wavelength, the trap region acts as an attractive potential for photons. If the lateral dimension of the trap region is comparable to the wavelength of the cavity-photons or polaritons, discrete 1D or 0D modes evolve in the system.

\begin{figure}[hb]
\centering
\includegraphics[width=8cm]{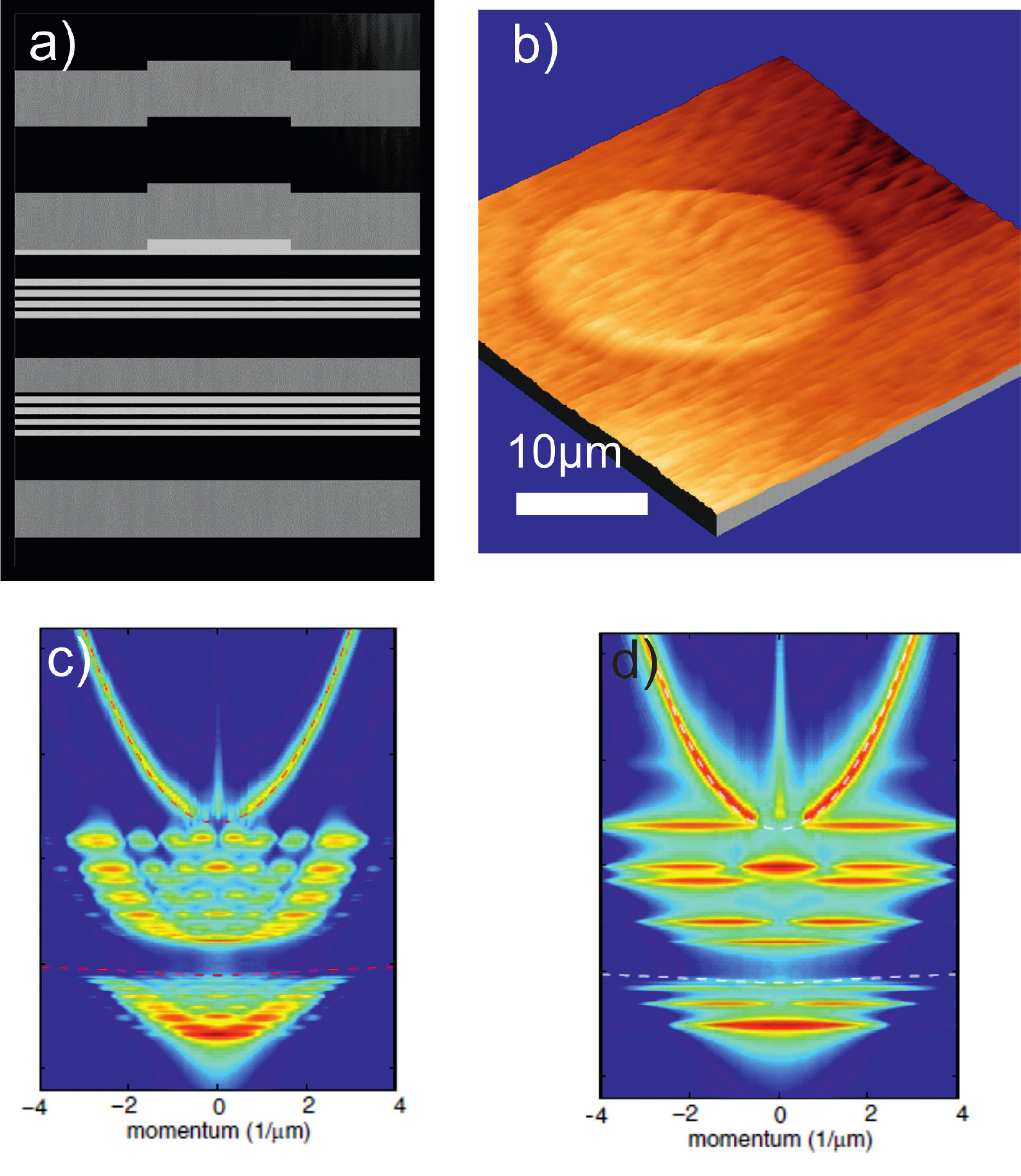}
\caption{\label{fig16} (a) Schematic drawing of the mesa trap structure to provide a lateral confinement for polariton condensates. (b) AFM image of the surface of an overgrown trap. Simulation of the energy-momentum dispersion of a trap with a diameter of 8.6 $\mu$m (c) and 3.6 $\mu$m (d) respectively. Reprinted from \cite{Kaitouni-PRB06},  Copyright (2006) by The American Physical Society. }
\end{figure}

Such a structure, which is sketched in Fig. \ref{fig16}a), can be realized e.g., via molecular beam epitaxy (MBE) growth in a three-step etch-and-overgrowth procedure: First the bottom DBR mirror and the cavity layer including all optical active regions (QWs) are grown. In case of a AlAs-cavity including GaAs-QWs, a thin GaAs-capping layer should be grown on top of the AlAs-cavity-layer to prevent the sample from  oxidation \cite{Winkler-NJP15}, and ideally be placed at a node of the optical field to circumvent absorption. This structure then gets transferred out of the MBE-reactor for patterning of the MC via lithography and etching prior to epitaxial overgrowth. Routinely, electron beam lithography is used to define the traps. However, advanced techniques to create three dimensional nanostructures \cite{mai20123d, Pires07052010} have a great potential to implement advanced geometries. 
In fact, the shallow elongation of the MC spacer is preserved even after deposition of the several microns thick top DBR mirror and can be fully resolved on the sample surface, e.g. via atomic force microscopy  (Fig. \ref{fig16}b).

A simulation of the polariton energy-momentum dispersion in such a mesa is shown in Fig.~\ref{fig16}c and d). Due to the finite size effect, a strong mode quantization is evident, which scales with the size of the lateral confinement. Emission from the planar microcavity persists in the background, however this emission can be significantly blue-shifted by increasing the etch depth.

The confinement of polaritons in such a structure can be rather accurately modeled by calculating the single particle energy levels in a finite, cylindrically symmetric (polariton) potential well.  The (lateral) well depth is given by the height of the mesa (and the according shift in the Bragg resonance energy).
Then, the Schr\"odinger equation can be solved for a particle with the polariton effective mass $m_{LP}$, yielding the discrete energy levels which can be seen in Fig. \ref{fig16} c) and d) for a diameter of 10 and 3.6 $\mu$m \cite{Kaitouni-PRB06}.
The low dimensional confinement, which can be designed in a wide range, can be utilized to engineer the scattering properties in such structures. A detailed discussion of the relaxation processed in such low-dimensional polariton structures is given in \cite{Paraiso-PRB09}. Noteworthy, it is found that the phonon relaxation between localized levels in the trap is enhanced (compared to 2D samples) as a consequence of momentum space selection rules, whereas energy transfers between 2D and 0D polaritons are strongly suppressed. Consequently, it is suggested that, under non-resonant excitation conditions, the 0D polariton states have to be directly populated from the exciton reservoir, rather than from the 2D polaritons.

The first demonstration of polariton condensation under non-resonant pumping in a mesa was discussed quite recently in \cite{Winkler-NJP15}, where the authors investigated a single trap with a diameter of 6 $\mu$m and a depth of $\sim 30$ nm.


\begin{figure}[hb]
\includegraphics[width=8cm]{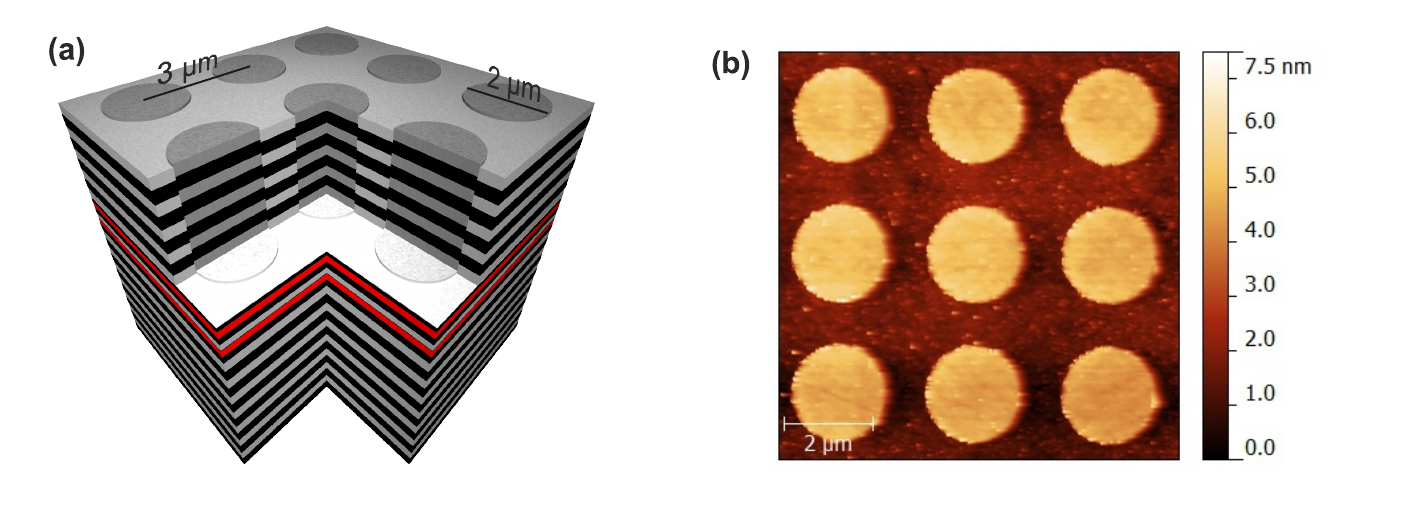}
\caption{\label{fig:fig4} Polaritons in a square lattice potential landscape. (a) Schematic drawing of the investigated square lattice structure consisting of buried polariton traps. (b) Atomic force microscopy image of the etched cavity surface, depicting the morphology of the mesa structure. }
\end{figure}

The full potential of this technique becomes obvious, when the traps are arranged in more complex geometries, such as square lattice arrangements (Fig.~\ref{fig:fig4}). The polariton wave functions can penetrate into the barrier leading to evanescent photonic coupling between neighboring sites which is accompanied by the formation of a band structure. Using the linear exciton-photon coupling Hamiltonian\cite{Deng2010}, it can be shown that the structure of the energy bands imposed on the cavity photon mode by the periodic potential translates into the band structure of the polariton spectrum \cite{Boiko2008} (see section \ref{bandstructure}).

\begin{figure}[htb]
\includegraphics[width=8cm]{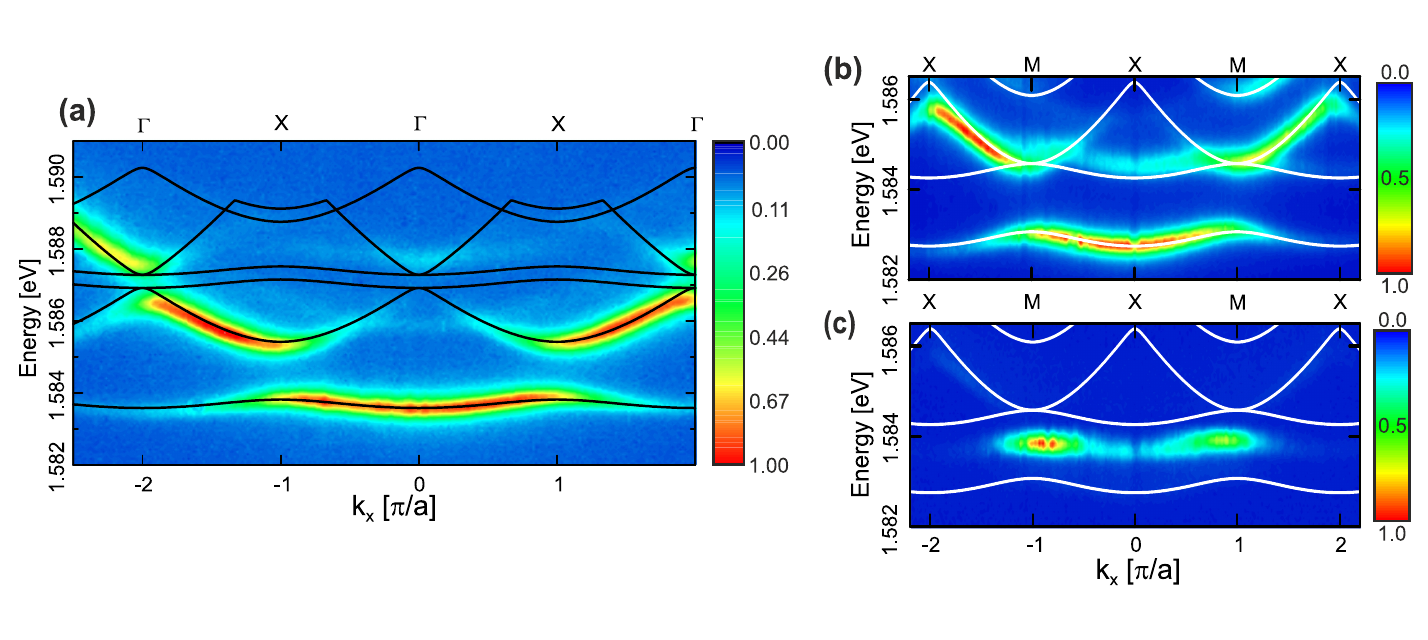}
\caption{\label{fig:fig5} Polariton condensation in a square lattice(a) Photoluminescence spectrum of a square lattice with a lattice constant a=3 $\mu$m. The axis shows the momentum in units of the BZ, while the upper axis marks the symmetry points (Fig 4a). The band structure calculated theoretically is plotted in black lines. (b) Zoom onto the band gap region of a square lattice with a lattice constant $a=3$ $\mu$m recorded along the X-M symmetry direction revealing a complete gap between the M and X symmetry points of the first BZ. (c) For large pump power (1.3 times power at the condensation threshold $P_{th}=5.7$ mW), polariton condensation in the full gap near the M-point is observed. The figure is reprinted with permission from \cite{Winkler-NJP15} }
\end{figure}

A comparison between experiment and theory is shown in Fig.\ref{fig:fig5}(a), recorded for a square lattice with a constant of 3 $\mu$m below condensation threshold from the $\Gamma$-point to the X-point of the BZ.  The wave-function overlap is sufficient to form distinct bands for the three lowest energy levels, which can be perfectly reproduced by a model with realistic parameters: $V_L=3.2$ meV, $m_{LP}=5.6\times10^{-5} m_e$.

The nonlinear characteristics of the polaritons in such a lattice is particularly manifested close to the M-point of the BZ, which is shown in the close up image of the gap between the s and p-band in our system, taken along the X-M symmetry direction below (Fig. \ref{fig:fig5}(b), $P \sim 0.2 P_{th}$) and above (Fig. \ref{fig:fig5}(c), $P \sim 1.3 P_{th}$) the condensation threshold. With increasing pump power, the condensate forms in the vicinity of M-points of the BZ blue shifts into the full gap of the linear spectrum. This behavior suggests the formation of a spatially localized two-dimensional gap soliton state \cite{Ostrovskaya2013}, also observed  in the potentials created by surface acoustic waves\cite{cerda-PRL13}. Although these gap states are analogous to their 1D counterparts observed in modulated photonic wires\cite{winkler2015collective}, the essential requirement for their existence in 2D is the complete gap which is only available at the M-point of the square lattice BZ.

\section{Conclusions and outlook}

Thanks to the rapid advances of the techniques which are required to manipulate light and matter waves on the scale of their characteristic wavelength, it has become possible to engineer the potential landscapes of light-matter particles in semiconductors. As a result Bose-Einstein condensation of exciton-polaritons in semiconductor systems has been demonstrated in a variety of trapping geometries.
The possibility to create polariton condensates in single traps, to engineer and manipulate scattering mechanisms, to control the propagation of polaritons on channels and to form versatile potential lattices with complete bandgaps and even Dirac-like dispersions opens the way towards more advanced fundamental studies and applications of polariton physics. One interesting opportunity, which we have touched in this article, can arise by utilizing polaritons in channel structures for integrated polaritonic schemes: While state-of-the-art microcavities facilitate polariton and soliton propagation over hundreds of microns close to the speed of light, the polaritons' matter part allows for efficient external manipulation via electro-optical and all-optical methods. Utilizing materials with large exciton binding energies, such as organic semiconductors \cite{plumhof2014room}, GaN\cite{christopoulos2007room}, ZnO \cite{lu2012room}, or sheet layers of transition metal dichalcogenides \cite{liu2015strong} can hence lead to novel architectures in integrated photonics based on coherent bosonic states. Another interesting option involves exploiting polaritonic effects to create photonic topological insulators \cite{Nalitov-PRL15, bardyn2014topological}, which facilitate unidirectional propagation of light in edge channels protected from back-scattering. While the investigation of topological effects in coupled light-matter systems is extremely appealing from a fundamental point of view, there is a whole zoo of possible applications arising from such effects. Further improvements of trapping schemes at the sub-micron scale could enable the regime of polariton blockade within reach, where non-linearities occur on the single polariton scale. Successful engineering of potentials allowing for single polariton localization is still a major challenge in the field, however it carries the promise  to open new directions in the research field of quantum polaritonics.

\subsection{Acknowledgment}

The authors thank A. Kavokin, I.A. Shelykh, I.A. Savenko, H. Flayac, T. Fink, T.C.H Liew, A. Nalitov, N. Gregersen, N.Y. Kim, S. Brodbeck, B. Bradel, A. Schade, A. L\"offler, A. Imamoglu, V.D. Kulakovskii, L. Worschech, A. Forchel for collaborations, uncountable discussions and support throughout all the years.
Expert technical support in sample preparation by M. Emmerling, A. Wolf, M. Wagenbrenner, S. Handel, S. Kuhn and T. Steinl is gratefully acknowledged. This work has been funded by the State of Bavaria, and the Australian Research Council. M.D. Fraser acknowledges funding by ImPACT Program of Council for Science, Technology and Innovation (Cabinet Office, Government of Japan).

\section{References}



\end{document}